\documentclass{PoS}

\def\lsim{\mathrel{\rlap{\lower4pt\hbox{$\sim$}}
    \raise1pt\hbox{$<$}}}                

\newcommand{\ra}        {\rightarrow}
\newcommand{\raa}       {\mbox{$\rightarrow$}}

\newcommand{\TeV}       {\mbox{TeV}}

\newcommand{\GeV}       {\mbox{GeV}}

\newcommand{\invfb}     {\mbox{fb$^{-1}$}}
\newcommand{\fb}        {\mbox{fb}}
\newcommand{\invpb}     {\mbox{pb$^{-1}$}}
\newcommand{\pb}        {\mbox{pb}}

\newcommand{\ee}        {\mbox{$e^+ e^-$}}
\newcommand{\gamgam}        {\mbox{$\gamma\gamma$}}

\newcommand{\met}    {\mbox{${\hbox{$E$\kern-0.6em\lower-.1ex\hbox{/}}}_T$}} 
\newcommand{\mht}    {\mbox{${\hbox{$H$\kern-0.75em\lower-.05ex\hbox{/}}}_T$}} 

\newcommand{\pt}    {\mbox{$p_T$}}

\newcommand{\abseta}    {\mbox{$|\eta|$}}

\newcommand{\ttbar}     {\mbox{$t\overline t$}}

\newcommand{\ppWHnbb}     {\mbox{$p\bar{p} \raa WH \raa \ell \kern-0.45em\lower-.05ex\hbox{/}\nu b\bar{b}$}}

\newcommand{\WHnbb}     {\mbox{$WH \raa \ell \kern-0.45em\lower-.05ex\hbox{/}\nu b\bar{b}$}}

\newcommand{\Zmumu} {\mbox{$Z \raa \mu \mu$}}

\def\bb{$b\bar{b}$}
\newcommand {\ppbar}         {\mbox{$p\bar{p}$}}

\def\lmet{$WH\rightarrow \ell\kern-0.45em\raise0.19ex\hbox{/} \nu b\bar{b}$}

\def\tprim{$t'$}
\def\bprim{$b'$}
\def\Wprim{$W'$}
\def\Zprim{$Z'$}

\newcommand {\zz}            {\mbox{$ZZ$}}
\newcommand {\mzz}           {\mbox{$M_{ZZ}$}}
\newcommand {\zzllll}        {\mbox{$ZZ \to \ell^+ \ell^- \ell^+ \ell^-$}}
\newcommand {\zzllnn}        {\mbox{$ZZ \to \ell^+ \ell^- \nu \nu$}}
\newcommand {\zzlljj}        {\mbox{$ZZ \to \ell^+ \ell^- j j$}}
\newcommand {\lljj}          {\mbox{$\ell\ell jj$}}
\newcommand {\llmet}         {\mbox{$\ell\ell + {\not \! E}_{T}$}}

\newcommand{\bs}{\ensuremath{B_s^0}}
\newcommand{\bd}{\ensuremath{B^0}}

\newcommand{\mm}{\ensuremath{\mu^{+}\mu^{-}}}

\newcommand{\bsmm}{\ensuremath{\bs\ra\mm}}
\newcommand{\bdmm}{\ensuremath{\bd\ra\mm}}

\newcommand{\DO}   {D\O\ Collaboration}
\newcommand{\CDF}  {CDF Collaboration}

\newcommand{\LHCb}  {LHCb Collaboration}

\newcommand{\CDFnote}[3]{CDF Note #1, {\it #2}, (#3)}

\newcommand{\ATLASconf}[3]{ATLAS-CONF-#1, {\it #2}, (#3)}

\newcommand{\prl}[3]{{\sl Phys.\ Rev.\ Lett.} {\bf #1}, #2 (#3)}

\newcommand{\plB}[3]{{\sl Phys.\ Lett.} {\bf B#1}, #2 (#3)}

\newcommand{\jhep}[3]{{\sl J.\ High Energy Phys.} {\bf #1}, #2 (#3)}

\newcommand{\prD}[3]{{\sl Phys.\ Rev.} {\bf D#1}, #2 (#3)}
\newcommand{\prDR}[3]{{\sl Phys.\ Rev.} {\bf D#1}, #2(R) (#3)}
\newcommand{\npB}[3]{{\sl Nucl.\ Phys.} {\bf B#1}, #2 (#3)}

\newcommand{\epjC}[3]{{\sl Eur.\ Phys.\ J.} {\bf C#1}, #2 (#3)}

\newcommand{\hepex}[1]{arXiv:hep-ex/#1}

\title{Legacy limits and hints of New Physics at the Tevatron}

\ShortTitle{Searches for BSM at the Tevatron}

\author{\speaker{Arnaud Duperrin}\\
        Aix-Marseille University, CPPM, IN2P3-CNRS, 13288 Marseille, France\\
        E-mail: \email{duperrin@cppm.in2p3.fr}}

\abstract{This paper reviews results of beyond-the-standard model searches at the Tevatron
presented in a plenary talk at the Europhysics Conference on High Energy Physics (EPS-HEP) in Grenoble.
Here I present a selection of results from the CDF and D\O\ experiments that were shown at the conference during the parallel sessions.
Much more is available and can be found at the experiment's web pages\thanks{CDF results: {\sf http://www-cdf.fnal.gov/physics/exotic/}}~\thanks{D\O\ results: \sf{http://www-d0.fnal.gov/Run2Physics/D0Summer2011.html}}. This proceeding
essentially focuses on searches in the dilepton, diboson, and \ttbar\ final states.
Throughout the paper, hints of new physics observed at the Tevatron are also briefly discussed.
}

\FullConference{The 2011 Europhysics Conference on High Energy Physics, EPS-HEP 2011,\\
		July 21-27, 2011\\
		Grenoble, Rh\^one-Alpes, France}

\begin{document}

\section{Introduction: what are we looking for ?}
A large dataset of more than 10~\invfb\ of integrated luminosity is accumulated by the Tevatron experiments.
This represents a lot of well understood data that increase the experiment's chances of seeing something interesting and sensitive to new phenomena.
Since we believe that new phenomena are likely to manifest themselves as an anomalous production rate of a combination of the known fundamental particles, CDF and D\O\ developed a strategy of looking at "everything" and "everywhere" with as many combinations as possible.

It is clear that particle physics is at a stage where there is no unique way forward. We do not know what this new physics is and thus precisely how to search for it. However, despite different theoretical inputs, many beyond-the-standard models end up looking at the same thing phenomenologically. The main question is therefore: how are we going to reveal some of these models from an experimental point of view?
Before providing some examples of experimental searches, one should note that even if we do not find anything, these detailed measurements of new phenomena are still very important to develop a deeper understanding of the standard model itself, as illustrated in many occasions throughout this paper.

\section{The Tevatron accelerator}
The Tevatron is widely recognized for numerous physics discoveries and for many technological breakthroughs and all the exiting set of results described here will not have been possible without such outstanding performances. Both CDF and D\O\ experiments deeply thank the accelerator division for the hard work and all the great achievements. After more than two decades of operation, the Tevatron proton-antiproton collider shutdown on September 30, 2011, marking the end of the facility's 28-year lifetime.
The experiments are also working hard to make the best use of the delivered luminosity in a timely fashion at a time when the LHC is also excellently performing. For EPS-HEP 2011, CDF and D\O\ analyzed up to 9~\invfb\ of the delivered luminosity.

\section{The CDF and D0 detectors}
One of the primary reasons for having more than one experiment studying the same kind of signal is that an unexpected finding in a particle physics experiment must be reproduced to rule out the possibility that it is either a statistical fluctuation or something other than new physics.
At Fermilab, there are two such experiments: CDF and D\O. Although there are recognizable differences, both are general purpose detectors remarkably similar in their design. For instance, both have silicon tracking to do b-tagging. The tracks are also reconstructed with an outer tracker: scintillating fibers for D\O\ and an open drift cell for CDF. The magnetic field used for the charged tracks \pt\ measurements is a bit stronger at D\O\ (1.9 Tesla) than at CDF (1.4 Tesla). However, CDF tracker radius is almost a factor of two of the one at D\O\ and makes up for this difference in magnetic field. D\O\ has a liquid argon calorimeter and a more traditional technology is used by CDF with a Pb/Cu/Scintillator calorimeter, leading to a jet energy scale and a resolution that is fairly similar between the two experiments (1-3\%). Finally, the drift/scintillator counters are used as muons detectors and cover the outside of the detectors (\abseta<1.4 for CDF and \abseta<2 for D\O).

\section{Modeling and data}~\label{sec:modeling}
The modeling of the data differs from one analysis to another. However, in most of the cases we use {\sc pythia}~\cite{ref:pythia} and {\sc alpgen}~\cite{ref:ALPGEN} as Monte Carlo  matrix element Leading Order ({\tt LO}) event generators with CTEQ5L or CTEQ6L~\cite{ref:pdf} leading-order parton distribution functions to produce the shapes that we expect. The {\tt QCD} multijet shape and normalization are usually obtained from the data. The normalization of the $Z$+jets and $W$+jets components are either based on the measured cross sections~\cite{ref:Zjets_CDF} or obtained from the application of next-to-leading order cross-sections (or {\tt NNLO}) predictions~\cite{ref:ZWjets_D0}. Cross sections of other standard model backgrounds like \ttbar, single top, and diboson samples are taken from ~\cite{ref:ttbar}, \cite{ref:singletop}, and \cite{ref:mcfm}, respectively.
Some analyses directly fit the data to evaluate the $Z/W+$jets contribution.

\begin{figure}
\centering
\includegraphics[scale=0.5]{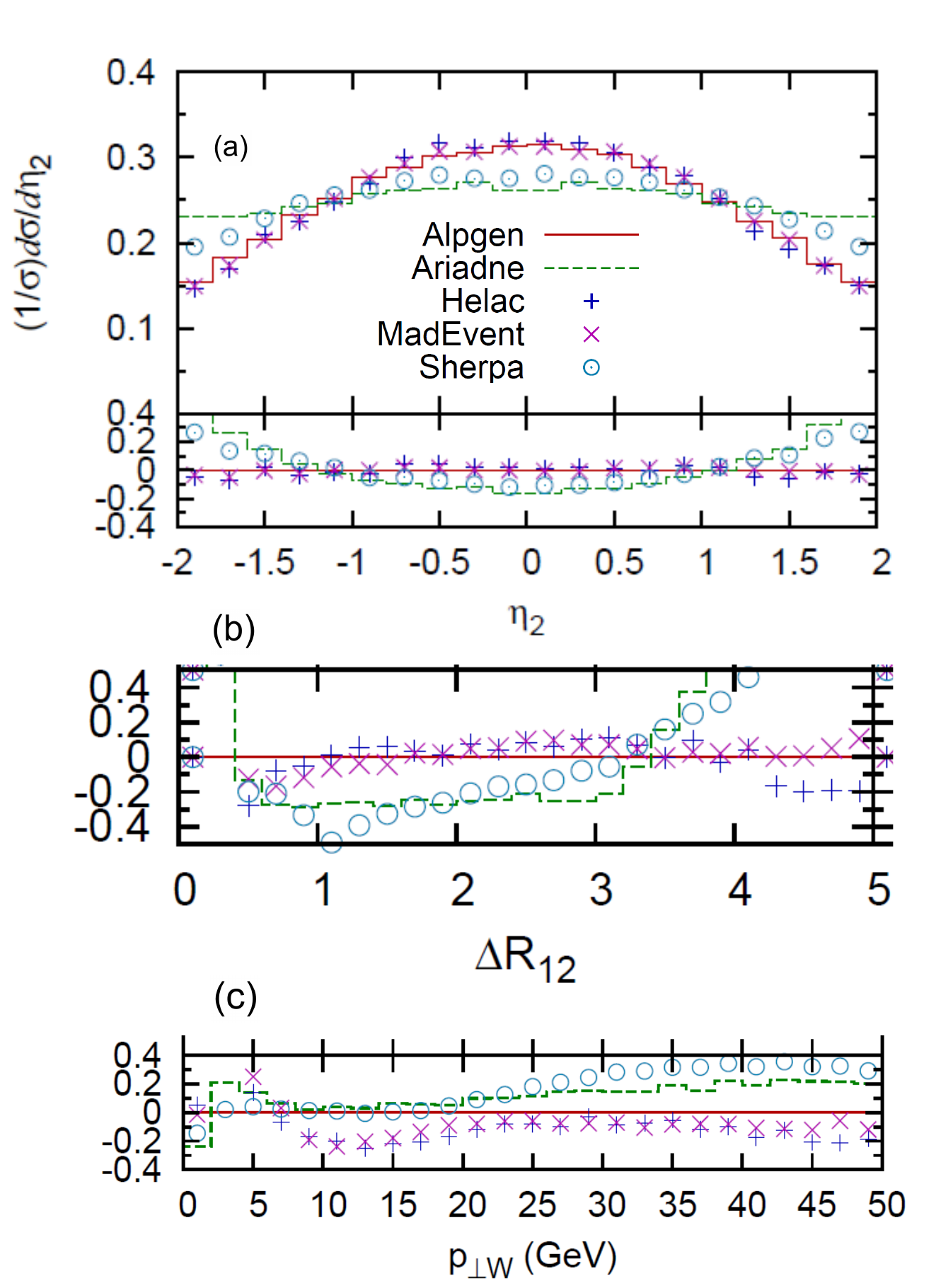}
\includegraphics[width=.5\linewidth,height=.6\linewidth]{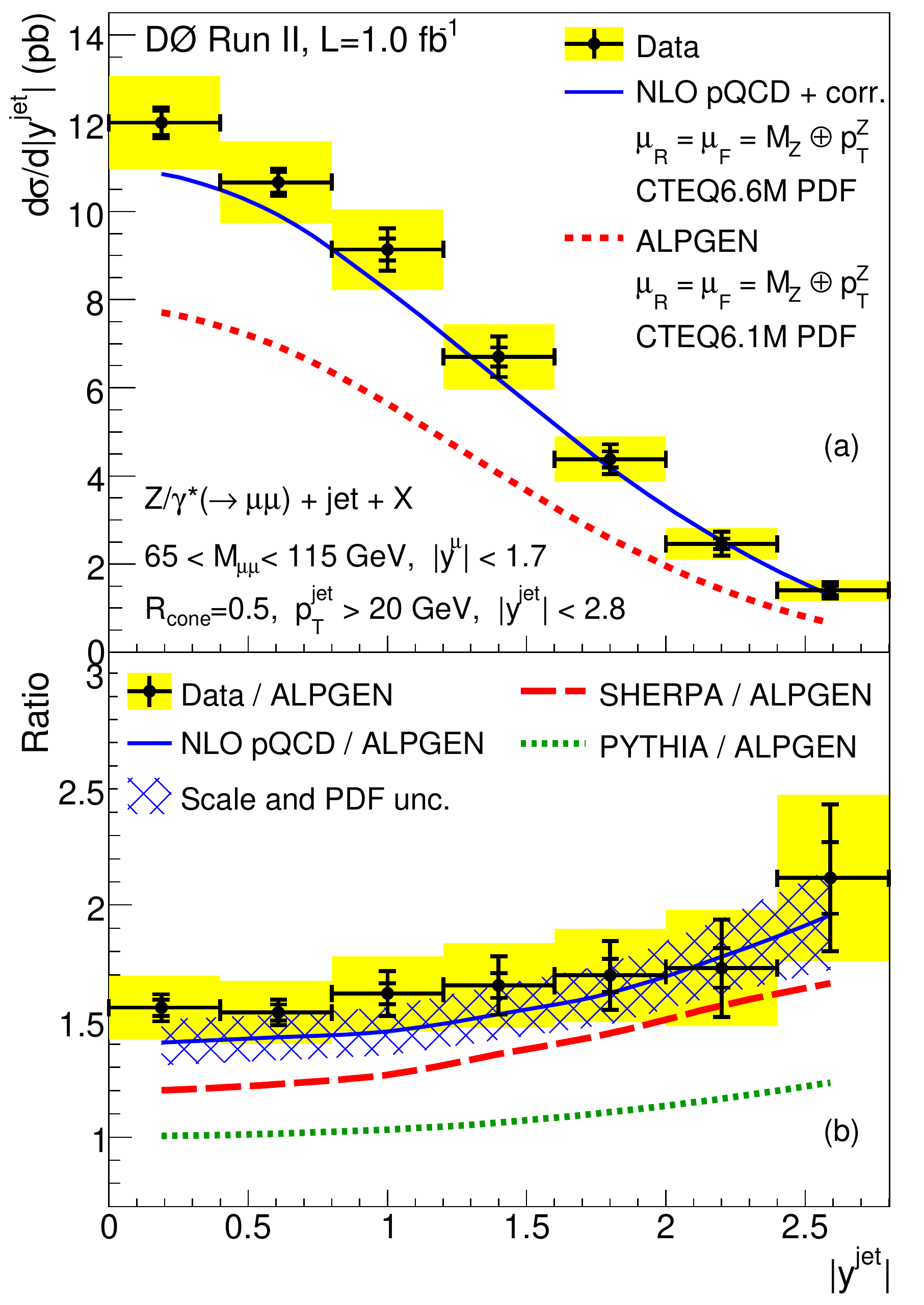}
\caption{Monte Carlo studies ({\it left plot}) in a sample of $W$+jets events at the Tevatron~\cite{ref:alpg_sherpa}. The
full line gives the {\sc alpgen} results, which serve as a reference and are compared to other generators. Figure (a) shows the predicted inclusive second leading jet $\eta_2$ normalized
to unit area. The relative differences with respect to the {\sc alpgen} results are represented in the lower in-sets plots. Figure (b) represents the separations between the jet 1 and 2 ($\Delta R_{12}$), and figure (c) the $W$ boson transverse momentum spectra. The measured cross section ({\it right plot}) by the D\O\ experiment is shown in bins of leading $|y_{\mbox{jet}}|$ for $Z/\gamma^*$+jets + X events~\cite{ref:D0_zjets_diff}. Figure (a) shows {\tt NLO} pQCD and {\sc alpgen} predictions and compared them to the data. Figure (b) displays the ratio of data and predictions from {\tt NLO} pQCD + corrections, {\sc sherpa}, and {\sc pythia} to the prediction from {\sc alpgen}.}
\label{fig:algen_sherpa_data}
\end{figure}

It is important to stress that if these simulation tools were substantially wrong, it is unlikely we would have been able to show the excellent agreement that have already been shown using the Tevatron data and published in hundred's of papers during Run II of the Tevatron.
However, it is also clear that doing an analysis is more complicated than just running a Monte Carlo tool out of the box because Monte Carlo's, although meant to reproduce the standard model, they remain approximations. There are many knobs to tune in order to actually reproduce what we see in our data and, at the end, different Monte Carlo's that are trying to reproduce the same things give different results.
While this has been known for a long time, it is useful to briefly remind a few general rules to provide a clear understanding of what we are doing at the Tevatron.

\begin{figure}
\centering
\includegraphics[width=.5\linewidth,height=.35\linewidth]{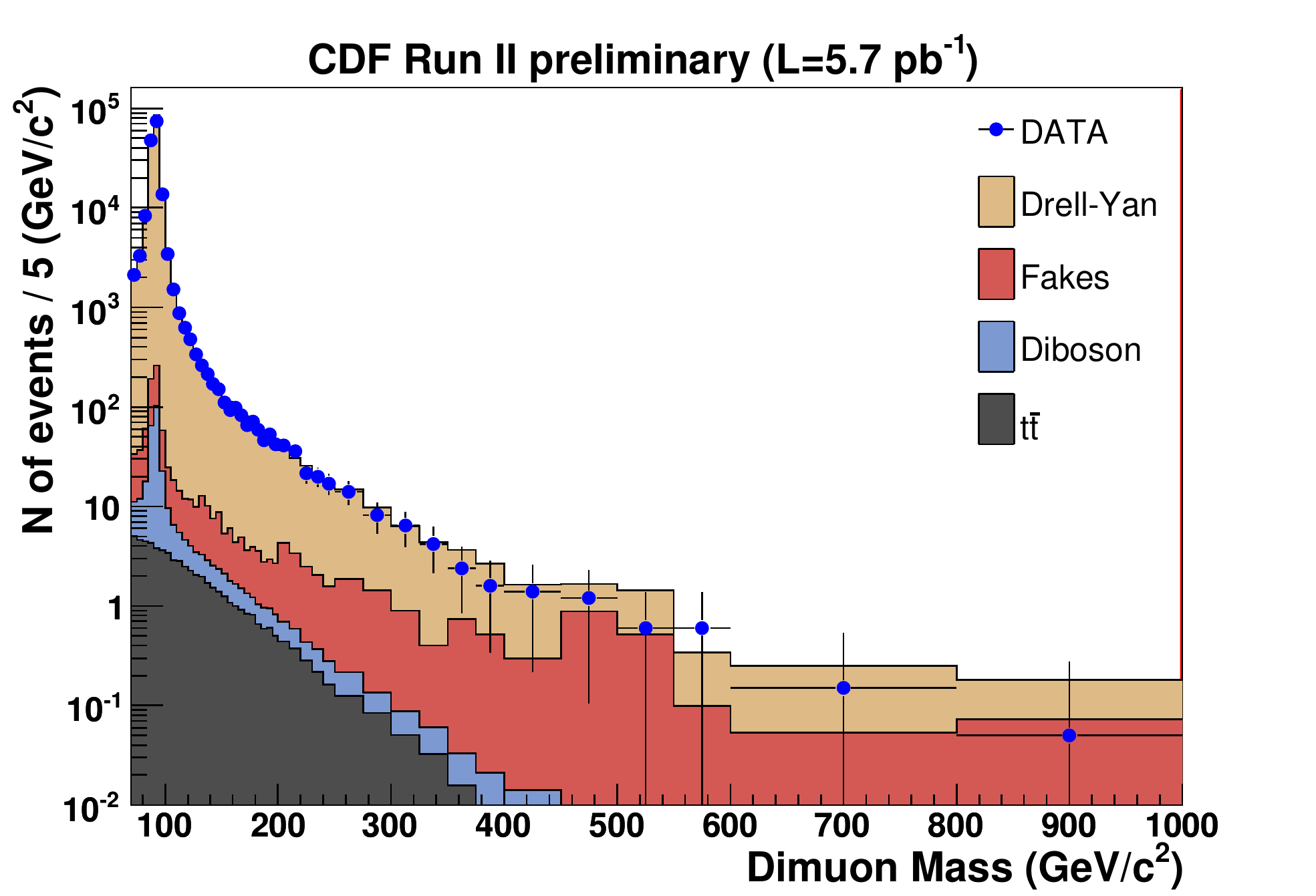}
\caption{CDF dimuon mass distribution measured in 5.7~\invfb\ of data~\cite{ref:CDF_RS_mumueegg}.}
\label{fig:CDF_mm}
\end{figure}

As mentioned above, Tevatron experiments use {\sc alpgen} for $W$ or $Z+$jets modeling. Figure~\ref{fig:algen_sherpa_data} on the left shows three experimental observable that are compared between 5 different fixed-order tree-level matrix element generators that use different development schemes of the hard partons into jets of hadrons~\cite{ref:alpg_sherpa}. Monte Carlo $W+$jets events are used for these studies and {\sc alpgen} serves as a reference when doing the ratio with the other generators. Note that the {\sc sherpa} version (v1.1) used in this study had a bug in the parton shower routine and {\sc sherpa} has since changed the shower treatment. More recent versions of {\sc sherpa} that can be found at the generator website~\cite{ref:sherpa_website} shows far fewer discrepancies. However, the essential features of the {\sc alpgen}/{\sc sherpa} differences are still present. The point to remember is that one can see clear differences between what {\sc alpgen} and {\sc sherpa} generators are predicting for the same standard model $W+$jets process. These published plots in Fig.~\ref{fig:algen_sherpa_data} also show comparisons between more generators.

Finally, it is even more important to compare these generators to what we see in our data (after correction of the reconstructed data distributions to particle-level distributions). Figure~\ref{fig:algen_sherpa_data} on the right shows a measurement of the differential cross section in $Z+$jets events from D\O\ for the jet rapidity at the particle level~\cite{ref:D0_zjets_diff}, together with the predictions from {\sc alpgen} and {\sc sherpa}.
In this specific example that addresses the modeling of variables dealing with angles, {\sc sherpa} appears here to be closer to the data. This is an interesting observation and the purpose of this discussion is to highlight that there are not only important differences between generators but there are also differences between these generators and our data. These two facts - very well known - imply that in order to do things correctly, experimentalists need to use data driven corrections to fix the modeling of these kinematic distributions provided by the Monte Carlo tools: mostly for jet eta, \mbox{$\Delta R$} between the jets, and the transverse momentum of the $W$ and $Z$ bosons (see for instance~\cite{ref:D0_ZptRW}).

Therefore, the point to be remembered is not which version of {\sc sherpa} or {\sc alpgen} has been used but rather that modeling uncertainties and data driven corrections are important issues when searching for new physics.

\section{Dilepton final states}

Extra Gauge bosons like \Zprim\ are predicted in various models such as the $E_6$ {\tt GUT}s models. The way experiments search for \Zprim\ is by reconstructing the dilepton mass as shown for instance in Fig.~\ref{fig:CDF_mm} using 5.7~\invfb\ of data in a CDF dimuon sample~\cite{ref:CDF_RS_mumueegg}. One can see very well the $Z$ mass peak and the Drell-Yan tail at high mass.
This old fashioned "bump hunt" takes advantage of precisely measured $Z$ production and decays, of lepton identification and trigger efficiencies high and very well understood, with low background and easily determined from fake {\tt QCD}.
By performing a scan of high mass resonances, Tevatron experiments set limits for various models, all of which predict different strength of the gauge coupling for this hypothetical \Zprim. Exclusion limits are therefore derived on the \Zprim\ mass as a function of the gauge coupling.
A limit at 1023~\GeV\ is set in the dielectron channel by the D\O\ experiment using a dataset of 5.4~\invfb\ assuming standard model like type for the \Zprim\ with somehow lower limit for $E_6$ \Zprim\ bosons~\cite{ref:D0_Zprim_ee}. The CDF result in the dimuon final state sets a Tevatron legacy limit at 1071~\GeV\ using 4.6~\invfb~\cite{ref:CDF_Zprim_mumu}.

The \Wprim\ can decay in the same way as the standard model $W$, with the exception that the $tb$ decay channel is accessible if the \Wprim\  is heavy enough. CDF published~\cite{ref:CDF_W_emu} a limit at 1.12~\TeV\ with a search for a \Wprim\ boson in the $e\nu$ decay mode, assuming the left-right symmetric model where the right-handed neutrino from the boson decay is light. To search for a \Wprim\ decaying into $tb$, D\O\ uses a similar analysis as the one for the single top measurement and sets limits on the left/right-handed couplings as function of the \Wprim\ mass~\cite{ref:D0_Wprim_tb}. A limit at 885 GeV is obtained for a purely right handed coupling of the \Wprim.

Table~\ref{tab:summary_Vprim} summarizes Tevatron mass limits at 95\% C.L. in searches for a resonance using
the sequential standard model \Wprim\ and \Zprim\ bosons. Limits available at the LHC are, however, already surpassing these results.

\begin{figure}
\centering
\includegraphics[width=.5\linewidth,height=.35\linewidth]{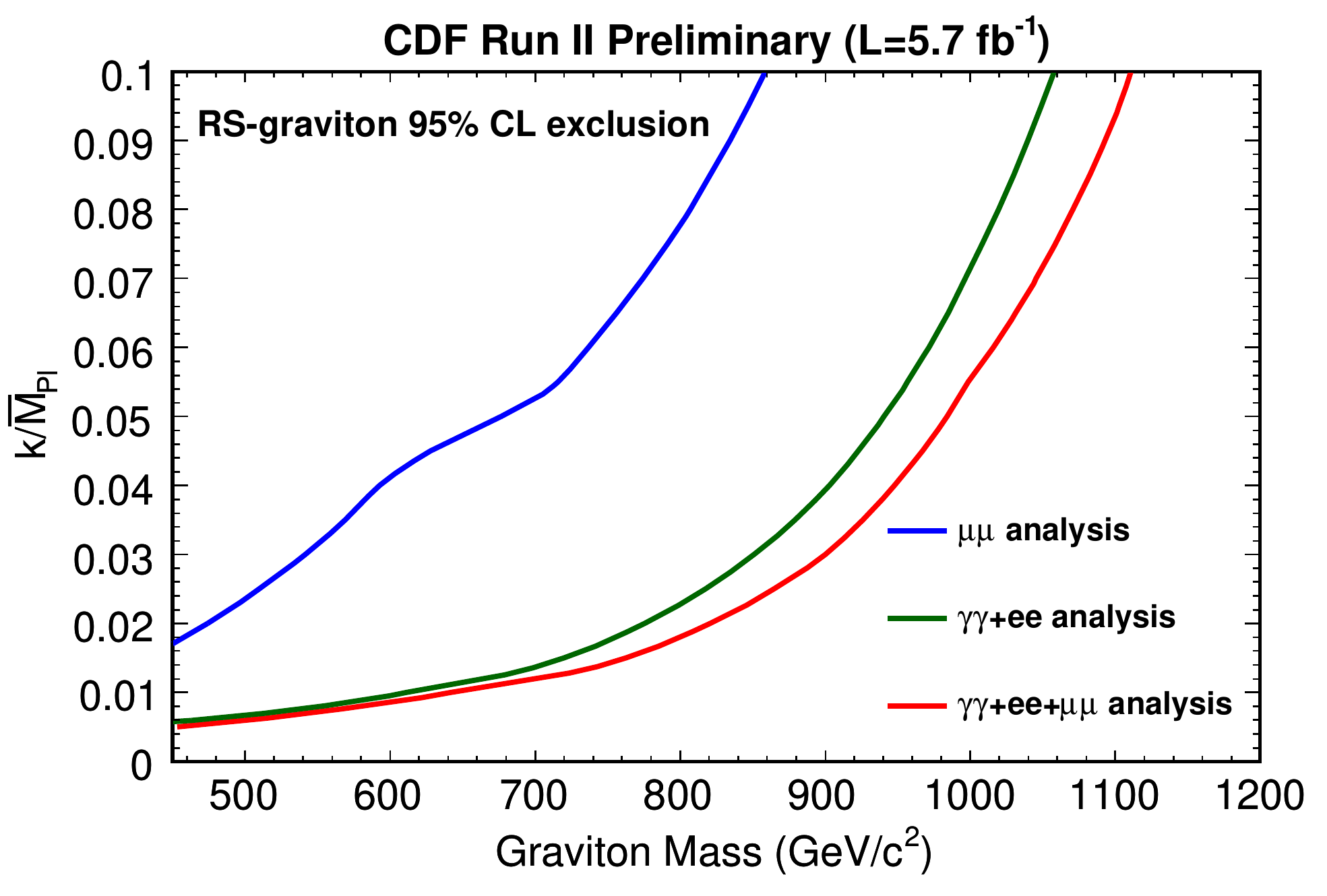}
\caption{CDF 95\% C.L. upper limit on $k/M_{\hbox{\rm Pl}}$
versus graviton mass from 5.7~\invfb\ of data for the $\gamma\gamma+ee+\mm$ final
states combined~\cite{ref:CDF_RS_mumueegg}.}
\label{fig:CDF_RS_ee_gg_mm}
\end{figure}

Models postulating the existence of large extra spatial
dimensions have been proposed to solve the hierarchy problem
posed by the large difference between the electroweak symmetry
breaking scale at 1~\TeV\ and the Planck scale, at which
gravity is expected to become strong. The first exited graviton mode predicted  by Randall and Sundrum ({\tt RS}) model could be resonantly produced at the Tevatron. The Graviton is expected to decay to fermion-antifermion and to diboson pairs. CDF and D\O\ searched for these resonances in their data.
Due to the Graviton having spin 2, the branching fraction to diphoton final state is expected to be twice the one in \ee\ final state.
In the dielectron and diphoton system, the background comes mainly from misidentified electromagnetic object and is estimated from the data. Combining \ee\ and \gamgam\ final states, limits could be set as a function of the Graviton mass and the coupling parameter ($k/M_{Pl}$) as shown in Fig.~\ref{fig:CDF_RS_ee_gg_mm}. For a value of 0.1 of this coupling, {\tt RS} graviton below 1~\TeV are excluded by CDF~\cite{ref:CDF_ZprimRS_ggee}.
One can further increase the sensitivity of the search by including the CDF dimuon final state~\cite{ref:CDF_RS_mumueegg}
and reach 1.1~\TeV, making this result the most stringent to date before EPS-HEP for {\tt RS} gravitons.

Table~\ref{tab:summary_RS} summarizes Tevatron graviton mass limits at 95\% C.L. using the Randall-Sundrum model.

\section{Diboson final states}

\begin{figure}
\centering
\includegraphics[scale=0.5]{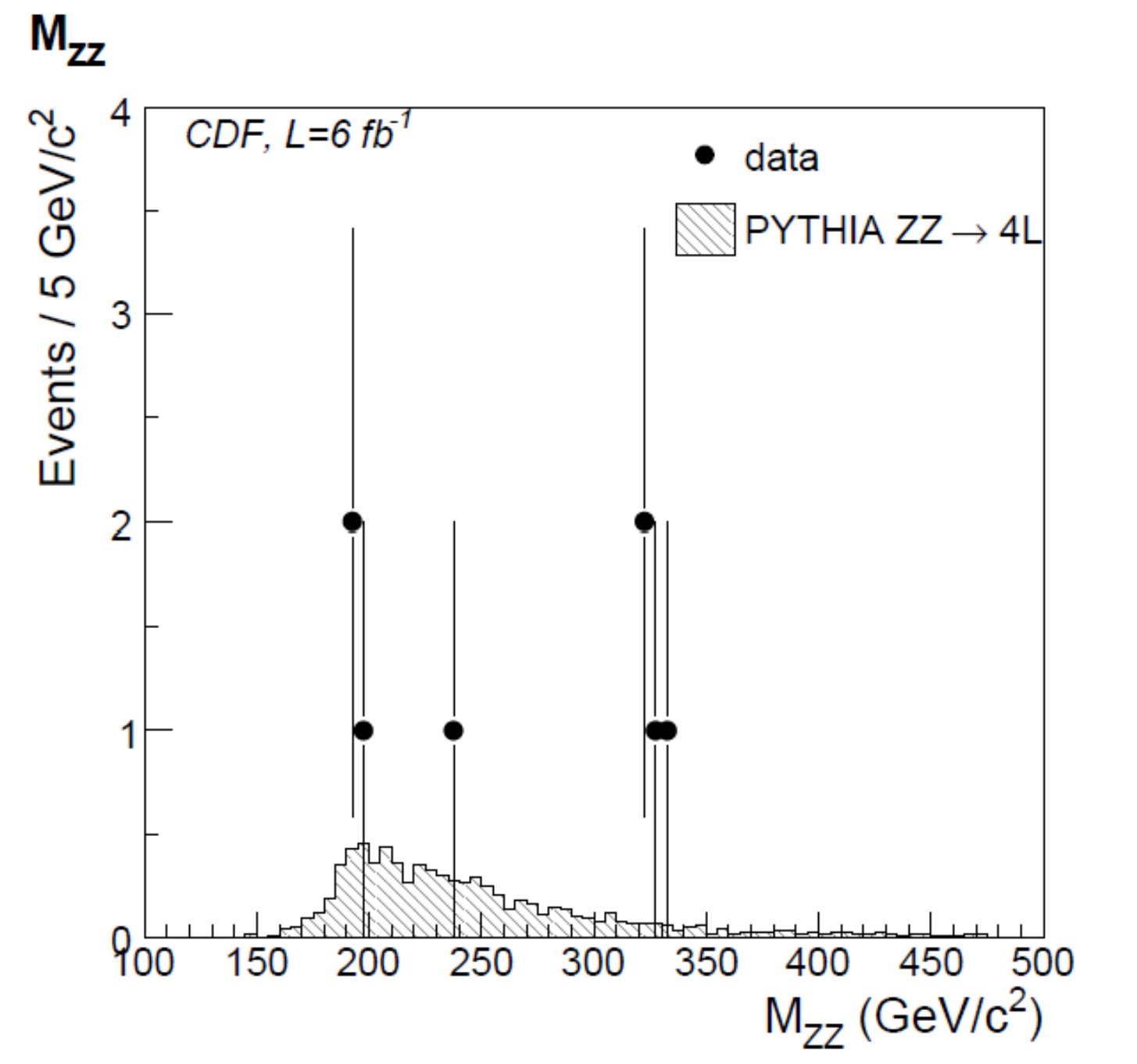}
\caption{Invariant masses \mzz\ for \zzllll\ candidates in 6~\invfb\ of data collected by the CDF experiment~\cite{ref:CDF_ZZ_4l_6fb}, along with the expected contribution from the standard model background as predicted by {\sc pythia}.}
\label{fig:CDF_ZZ_4l_6fb}
\end{figure}

The production of pairs of vector bosons is also a very interesting process because it allows to search for signal of new physics.
While it is already an extremely rare process in the standard model, Fermilab has gone down each step and measured each of the challenging cross sections. First measurements of $WZ$ and $ZZ$ where one of the $Z$ goes into \bb\ have been reported during the EPS-HEP conference. In many beyond-the-standard models scenarios, similar trends are repeating. For instance, dibosons are the dominant channels for high mass Higgs searches, and new physics scenarios predict particles such as {\tt RS} gravitons or \Wprim\ which would decay into diboson. Finally one could have final states where a $W$ is produced in association with a \Zprim\ decaying into a pair of jet, leading to an exotic $W$+2 jets final state.

If the \Wprim\  is heavy enough, it could decays in the $WZ$ final state. In a search with up to 5.4~\invfb\ of integrated luminosity, D\O\ exploits the property that the $W$ and $Z$ bosons result from the decay of a massive resonance and are therefore highly boosted~\cite{ref:D0_reso_WW_ZZ}.
Given the extended size of jets, the two jets from the hadronic decay of $W$ or $Z$ bosons with sufficient transverse momentum may be merged in a single jet whose mass corresponds to the original boson mass. Accepting single jet events significantly improves the sensitivity at high mass resonances.
D\O\ uses a sequential standard model  \Wprim\  boson as benchmark, combines 3 independent searches (in the fully leptonic, lepton +jets and dilepton final states), and excludes such \Wprim\ up to 690~\GeV.

One should also consider the case where a graviton decays to photons, leptons, and light jets can be suppressed, and dibosons become a discovery channel.
Both CDF and D\O\ searched for it and set limits which are given in table~\ref{tab:summary_RS} using a Randall-Sundrum as a benchmark model. The D\O\ search sets limits at 754~\GeV\ in $WW$ and 491~\GeV\ in the $ZZ$ finale state~\cite{ref:D0_reso_WW_ZZ}.

CDF presented updated results of a search for high-mass resonances decaying to $ZZ$ using 6~\invfb\ of data~\cite{ref:CDF_ZZ_4l_6fb}.
The search is performed in three distinct final states: \zzllll, \zzllnn, and \zzlljj. In the \zzllll\ channel, four events are consistent with a potential new resonance with mean value at 327~\GeV\ as shown in Fig.~\ref{fig:CDF_ZZ_4l_6fb}. Two of those events have unusually high $p_T(\zz)$. However, after combining with more sensitive searches in the \llmet\ and \lljj\ final states which did not show indication of a new heavy particle decaying to two $Z$ bosons, CDF sets 95\% C.L. upper limits on the production cross section times branching ratio to \zz\ for a Randall-Sundrum graviton $G^*$, $\sigma(\ppbar\to G^* \to\zz)$, which vary between 0.26~\pb\ and 0.045~\pb\ in the mass range $300<M_{G^*}<1000~\GeV$.

\section{A bump in CDF $W+$jets data}

\begin{figure}
\centering
\includegraphics[scale=0.33]{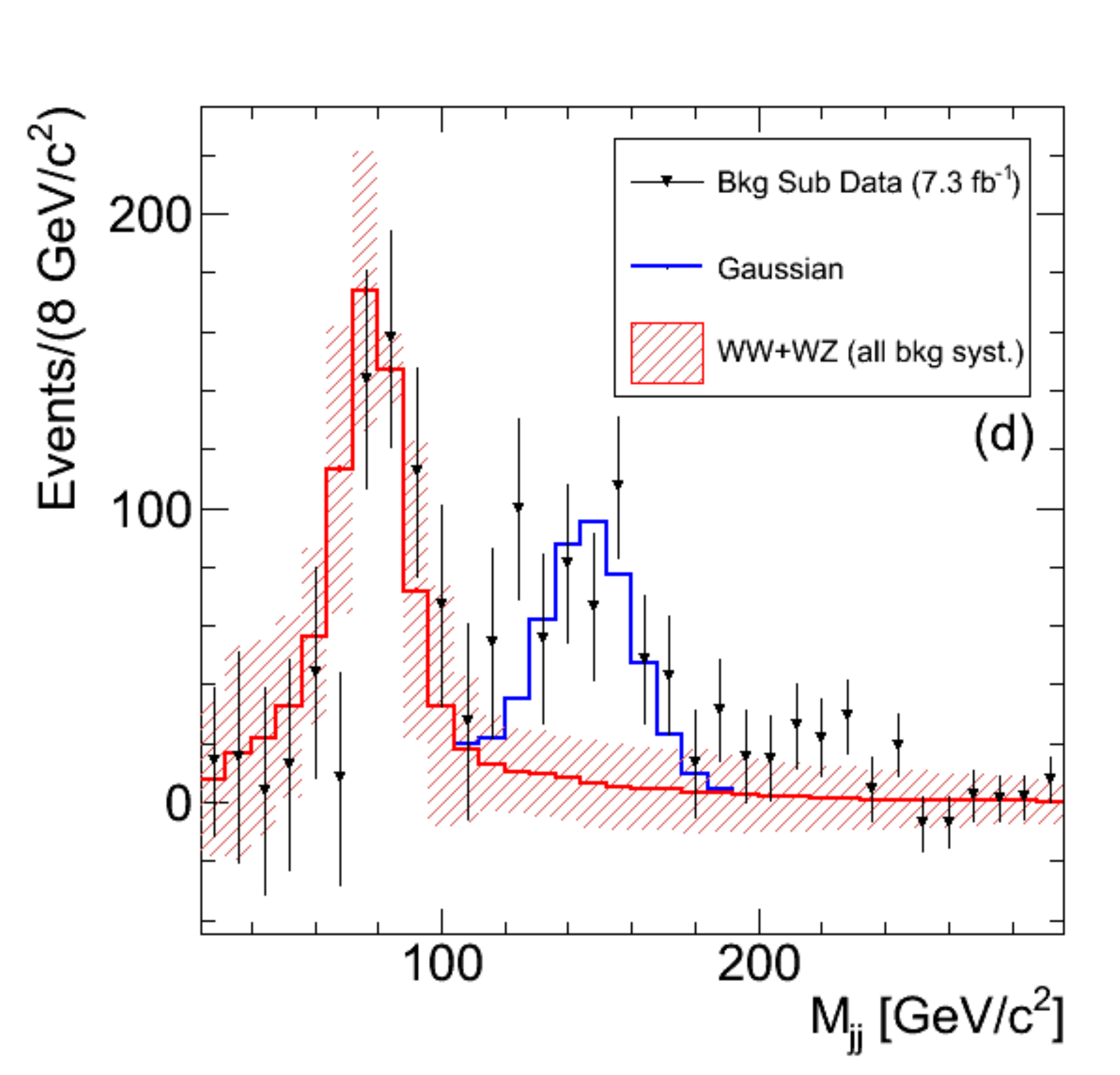}
\includegraphics[width=0.45\linewidth,height=0.4\linewidth]{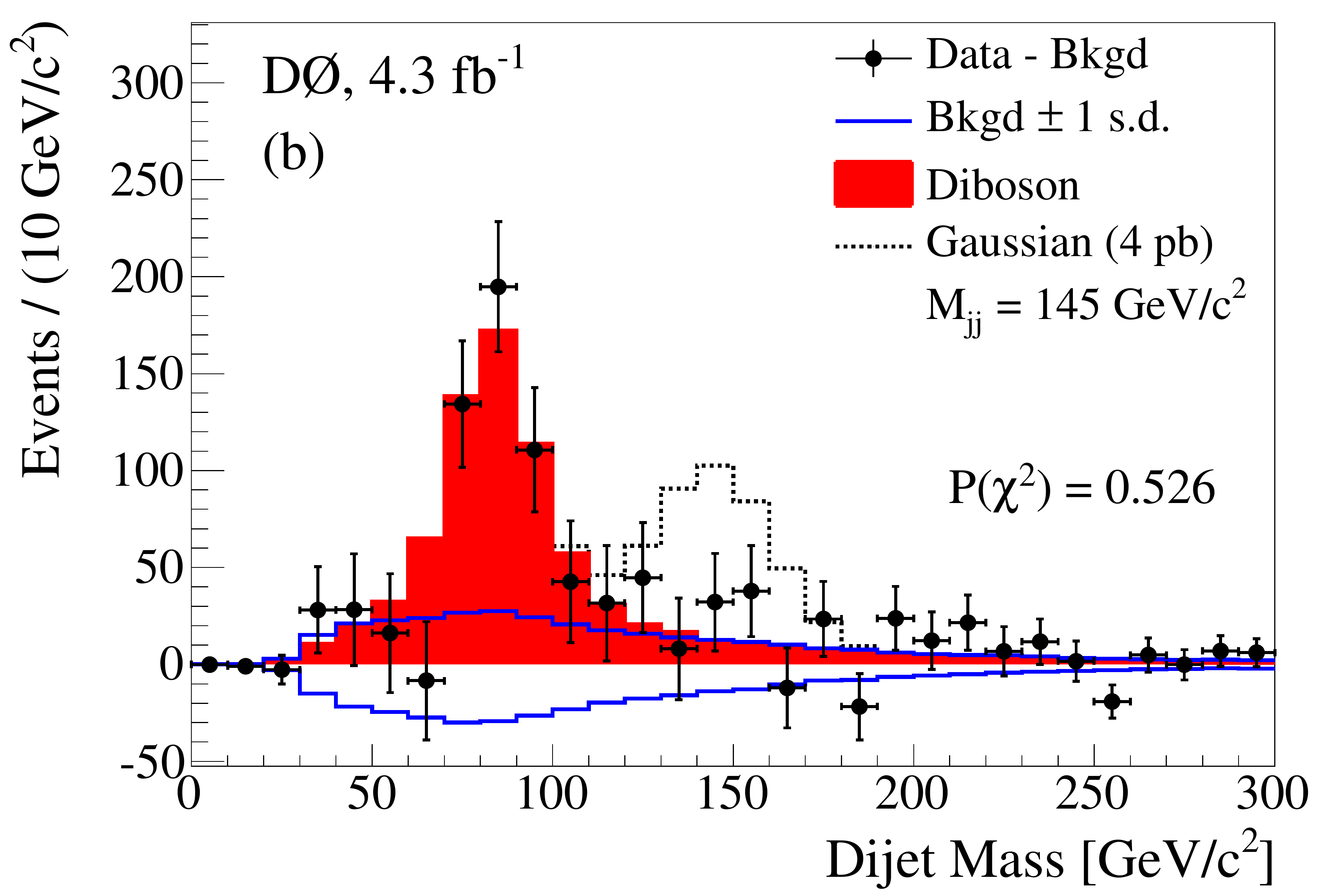}
\caption{CDF ({\it left}, \cite{CDF_Wjj_43_bump}, \cite{CDF_Wjj_73_bump}) and D\O\ ({\it right}, \cite{D0_Wjj_43_bump}) distributions of dijet invariant mass distribution for candidate events selected in an analysis of $W$+2 jet events after subtracting off the background except diboson. The black points represent the data. In the CDF figure, the red line plots the expected standard model background shape based on Monte Carlo modeling, the red shading shows the systematic and statistical uncertainty on this background shape. The blue histogram is the Gaussian fit to the unexpected peak centered at 144~\GeV. The D\O\ figure shows the relative size and shape for a model with a Gaussian resonance with a production cross section of 4~\pb. As is evident from this picture, no bump is observed that would correspond to the one seen by CDF.}
\label{fig:CDFD0Wjjbump}
\end{figure}

A couple of months ago, using 4.3~\invfb\ of data, CDF reported an excess of events when looking at $W$ bosons associated with 2 jets for a dijet mass around 145~\GeV~\cite{CDF_Wjj_43_bump}. This raised immediately interest and expectations turned to D\O\ for a potential confirmation of the observed bump.

The CDF selection includes one high \pt\ central and isolated electrons or muon above 20~\GeV, exactly 2 jets greater than 30~\GeV\ each and a cut on the transverse mass above 30~\GeV\ makes sure real $W$ events are selected.
The dijet invariant mass is required to be greater than 30~\GeV\ to select high mass dijet events and to shift them away from where the standard model diboson background is measured.
The dijet mass spectrum from CDF after subtracting off the background except diboson is shown in Fig.~\ref{fig:CDFD0Wjjbump}. The diboson peak is clearly seen as well as the excess above it. This excess is modeled by a gaussian and a width driven by the detector resolution.
The CDF analysis now uses 7.3~\invfb\ of integrated luminosity~\cite{CDF_Wjj_73_bump} and the significance of the excess is 4.1 sigma above the standard model prediction for a corresponding cross-section of about 4~\pb. In order to test the {\tt NNLO} contributions to the $W+$2 partons prediction, CDF compares {\sc alpgen} interfaced to {\sc pythia} for showering to a sample of $W+$2 partons simulated using {\tt MCFM} and extract a correction which is applied to the dijet invariant mass. This procedure returns a statistical significance of 3.4 sigma.

\begin{figure}
\centering
\includegraphics[width=0.45\linewidth,height=0.4\linewidth]{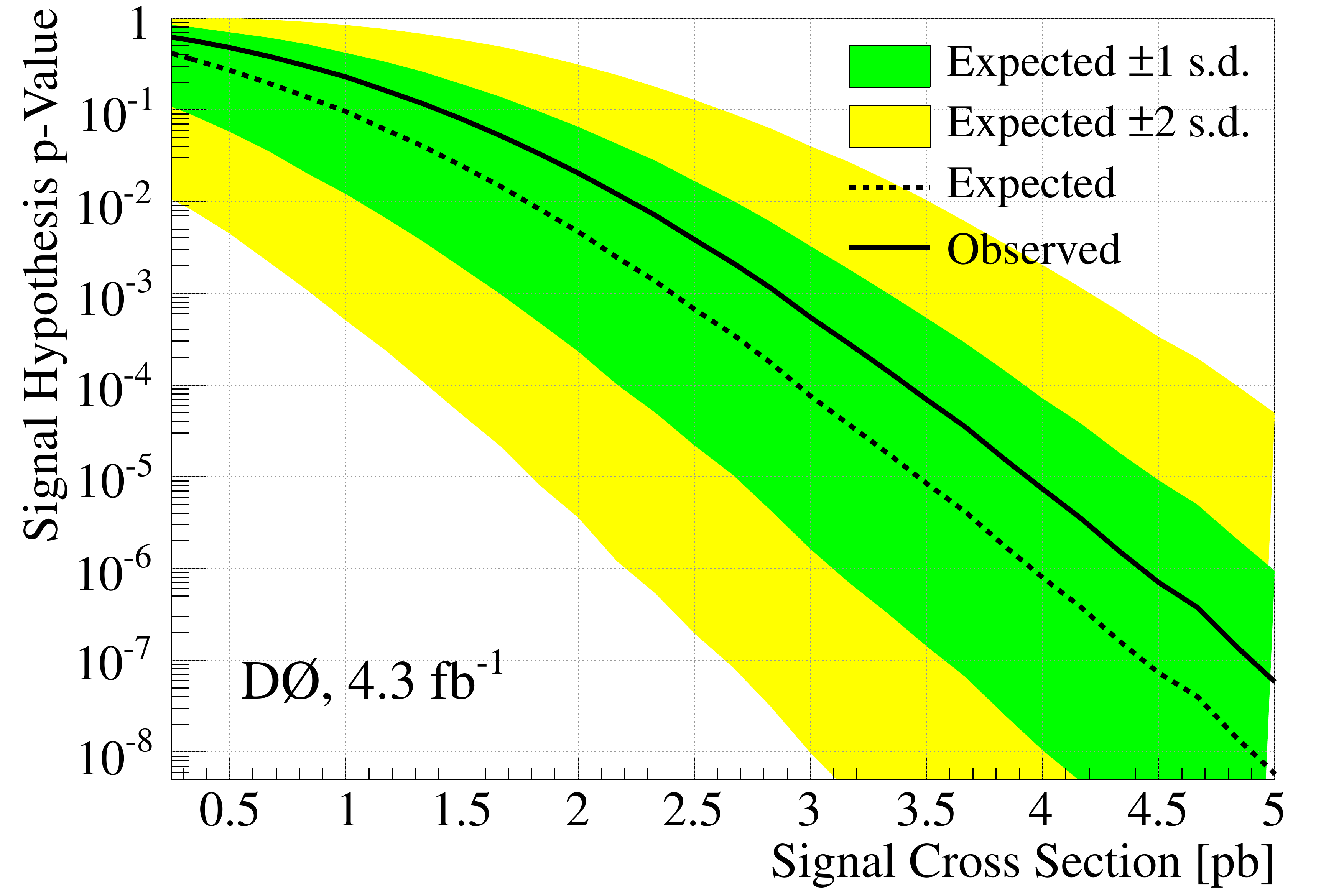}
\includegraphics[width=0.45\linewidth,height=0.4\linewidth]{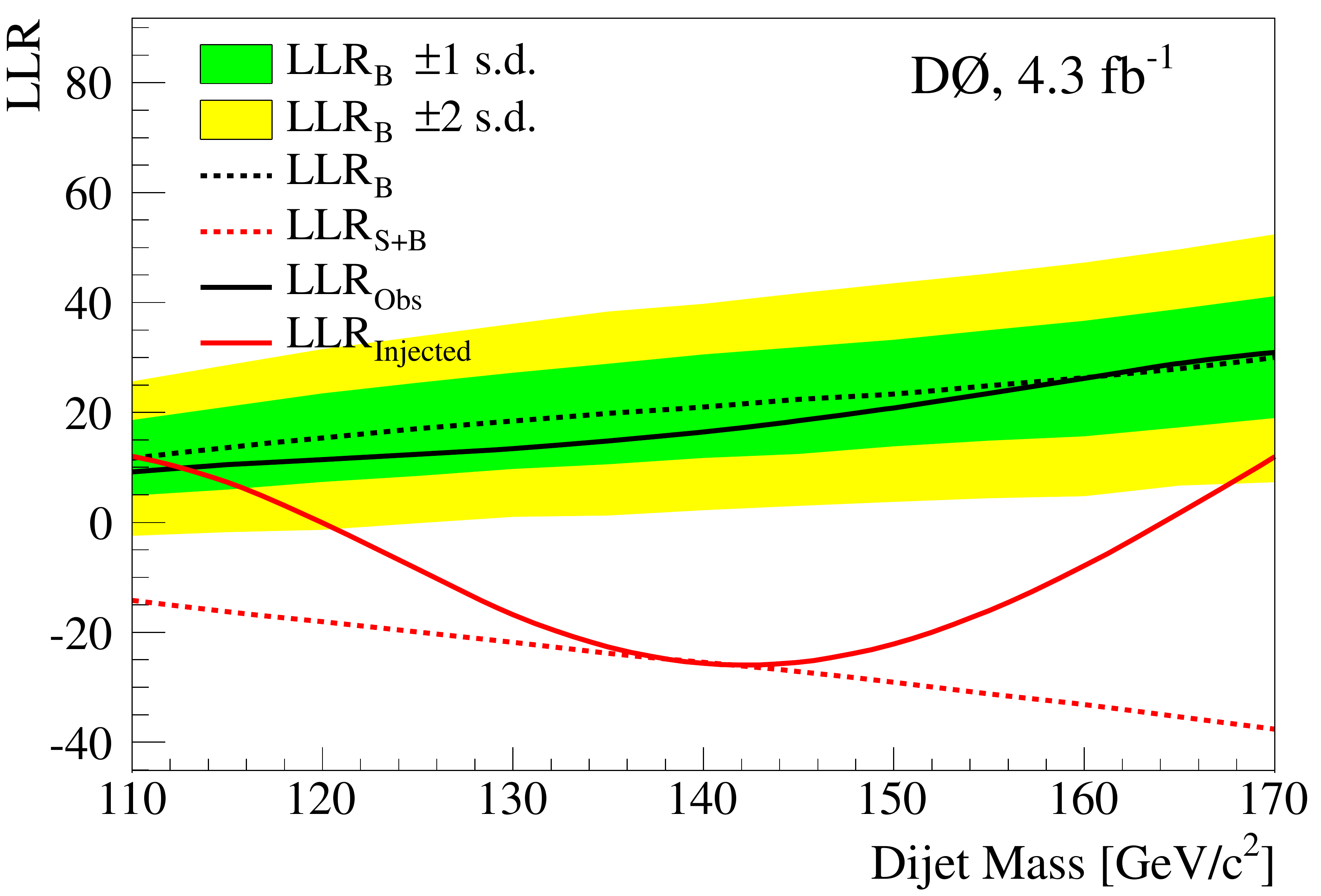}
\caption{Distribution of p-values ({\it left}) for the signal+background hypothesis with a Gaussian signal with mean of $M_jj = 145~\GeV$ as a function of hypothetical signal cross section (in \pb) in the D\O\ study of the dijet events selected in an analysis of $W$+2 jet events~\cite{D0_Wjj_43_bump}. Log-likelihood ratio test statistic ({\it right}) as a function of probed dijet mass (see text for more details).}
\label{fig:D0Wjjbump_pvalue}
\end{figure}

The D\O\ publication~\cite{D0_Wjj_43_bump} started with the on going diboson analysis and paralleled what CDF did. The D\O\ result after background subtraction is shown in Fig.~\ref{fig:CDFD0Wjjbump}. The data basically follow the diboson peak with no excess above it. The doted line represents the size of the signal that one would expect consistent with the CDF excess and the conclusion is that there is no evidence for such an excess in the D\O\ data. A particle with a cross section of 4~\pb\ is ruled out at the 4.3 sigma level. Since the cross section measured by CDF is of the "order of 4~\pb\ cross section", D\O\ provides figure~\ref{fig:D0Wjjbump_pvalue} allowing anyone to pick up whatever preferred cross section and derive its preferred number of sigma. Actually, CDF presented at the EPS-HEP conference a revised lower value for the cross section of $3.0 \pm 0.7 \pb$ using a procedure closer to what D\O\ did (see the reference~\cite{CDF_Wjj_73_bump} for all the details on this update).

Could D\O\ have missed such a structure in their data? To study this question, D\O\ uses a statistical test, shown in Fig.~\ref{fig:D0Wjjbump_pvalue}, which is the Log Likelihood Ratio ({\tt LLR}) of the S+B fit (signal+background) and B only fit where one can construct a probability from the ensemble test. The black doted line shows the background hypothesis only with its 1 and 2 sigma bands in yellow and green respectively. The solid black corresponds to the data. The red dotted line is the {\tt LLR} value for the S+B model with the 4~\pb\ signal. One can see that the data follow the background within less than one standard deviation and that such a signal hypothesis is clearly inconsistent with the data. The red solid line corresponds to the data where a 4~\pb\ signal is injected. It is pretty clear that in this case D\O\ would be sensitive to such a signal.

The fitted $W+$jets normalization is also shown to be consistent with the theory expectation. One of the pragmatic reasons for floating the cross section in the fit is because the $W+$jets is notoriously difficult to predict and also the dominant contributions. Therefore one can take the most conservative approach for letting the data determine the normalization. A fit to the D\O\ data of the standard model predictions plus a Gaussian signal template with $M_{jj} = 145~\GeV$ gives yields to a cross section of $\sigma(WX)\times B(X\rightarrow jj)= 0.82^{+0.83}_{-0.82} \pb$ or $\sigma(WX)\times B(X\rightarrow jj)= 0.42^{+0.76}_{-0.42} \pb$ when the diboson cross section is fixed to the standard model prediction. The fit is performed in the same way as in the $WZ$ search except that it now also includes the Gaussian signal template with a freely floating normalization. Fitted cross sections are both consistent with zero.

Should we also worry on what the sort of kinematic corrections applied to the simulation as then ones described in section ~\ref{sec:modeling} are doing? Could we hide an hypothetical signal? First of all, CDF and D\O\ include uncertainties on the modeling of the {\sc alpgen} variables described in section~\ref{sec:modeling}. For instance the D\O\ analysis considers systematic uncertainties on both the normalization and the shape of dijet invariant mass distributions. Therefore, there are systematics on these kinematic corrections as well as on the choice of renormalization and factorization scales, and on the {\sc alpgen} parton-jet matching algorithm. Finally, to make sure these corrections would not hide an hypothetical signal, D\O\ performs in addition an analysis without applying them. The relative change is not found to be large but when the corrections are applied, it improves the modeling substantially with the higher $\chi^2$ probability increasing from 0.53 to 0.74. There are indeed very good justifications to apply these corrections in some cases as already reported many times in this proceeding.

At the conference, CDF stated that data in the two experiments are compatible with each other within about 2 standard deviations after taking into account the systematic and statistical uncertainties~\cite{CDF_Wjj_73_bump}. Note that ATLAS released a study based on 1~\invfb\ where the measured dijet mass spectrum shows no significant excess over the standard model expectation~\cite{ATLAS_Wjj_1fb_bump}. A joint CDF/D\O\ task force has been formed to reconcile the two results.

\section{Leptoquarks}
Connecting the quarks and leptons is predicted in many extensions of the standard model, such as SU(5) grand unification, superstring, and compositeness models. It is actually interesting to see that the Tevatron is able still to set competitive limits on such leptoquarks ($LQ$). In this new result based on 5.4~\invfb~\cite{ref:D0_LQ}, D\O\ considers the case of first generation pair production of scalar leptoquarks in which one leptoquark decays to an electron ($e$) and a quark ($q$) and the other one to a neutrino ($\nu$) and a quark. The shape of the scalar sum of the \pt\ of the lepton, the missing transverse energy from the $\nu$, and of the two jets is used as a discriminant variable to set limits on the $LQ$ masses as function of the branching ratio $\beta$ of the leptoquark into $eq$. These limits are still slightly more constraining than those at LHC (with 40~\invpb) for beta below 0.35.

\begin{figure}
\centering
\includegraphics[width=0.65\linewidth,height=0.48\linewidth]{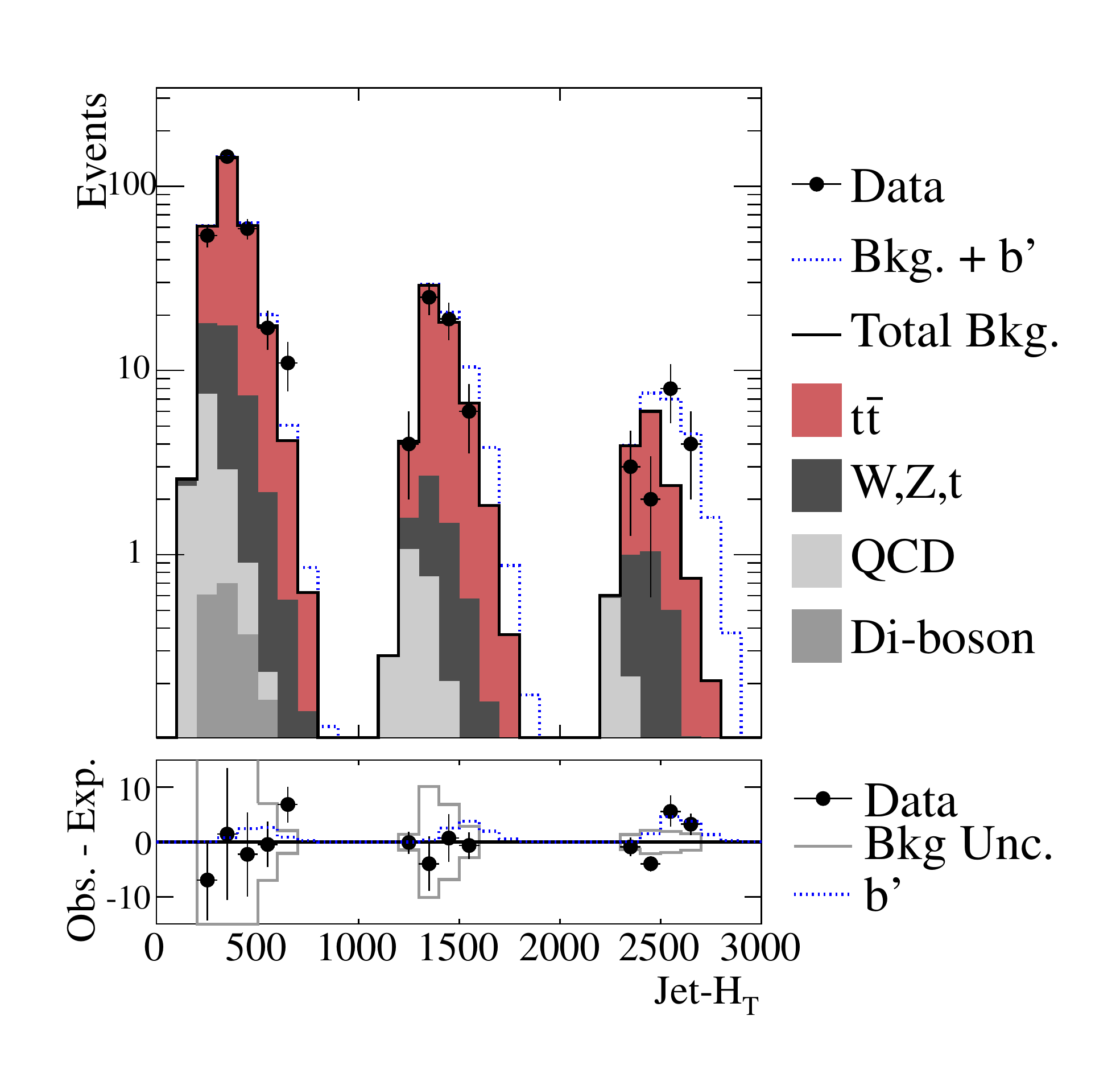}
\caption{CDF Jet-$HT$ distributions measured in 4.8~\invfb\ of data in a search for pair production of fourth-generation bottom-like chiral quarks ($b'$) each decaying promptly to $tW$~\cite{CDF_bprim}.}
\label{fig:CDF_bprim}
\end{figure}

\section{Top and anti-top final states}
The top quark is unique because of its large mass near 173~\GeV, which distinguishes it from the other fermions of the standard model.
Another very interesting way to search for new physics is therefore the \ttbar\ final state. The results presented in this plenary talk focus on $4^{\rm th}$ generation models and resonances in \ttbar.

\subsection{Fourth generation of fermions}
Measurements of the partial width of the $Z$ boson to invisible final states at LEP do not exclude the existence of a fourth generation of fermions as long as its neutrino is more massive than 45~\GeV. If the new physics is an additional sequential fourth generation, there would be two new heavy quarks that we call \tprim\ and \bprim\ for the up-type and down-type partner respectively. An important characteristic of this new quark doublet is that the mass splitting between the \tprim\ and the \bprim\ quarks is constrained by the electroweak precision tests to be small~\cite{ref:Gfitter}.
A heavy $4^{\rm th}$ family could also naturally play a role in the dynamical breaking of electroweak symmetry and the concept of an elementary Higgs scalar field would no longer be appropriate. Finally, a $4^{\rm th}$ family might also solve baryogenesis related problems, by visible increase of the measure of CP violation.

The CDF experiment searched for pair-production of a \bprim\ followed by prompt decay to a top quark and a $W$ boson assuming a 100\% branching ratio~\cite{CDF_bprim}. The assumption that \bprim\ decays exclusively to $tW$ is reasonable if the coupling to light quarks is small, as expected from precision meson-mixing measurements. This analysis uses the \ttbar\ lepton+jet final state where the standard model backgrounds can be separated from a potential signal by comparing the total reconstructed transverse momentum ($HT$) in the event for different jet multiplicity as represented in Fig.~\ref{fig:CDF_bprim}. The dominant background remains the \ttbar\ final state. The jet-$HT$ distribution is used to derive the cross-section limit as function of the \bprim\ mass. Though there are events with larger $HT$ than expected in the 7-jet event distribution, CDF does not see evidence of a signal and set the most restrictive direct lower limit on the mass of a down type fourth-generation quark \bprim\ at 372~\GeV.

\begin{figure}
\centering
\includegraphics[width=0.6\linewidth,height=0.48\linewidth]{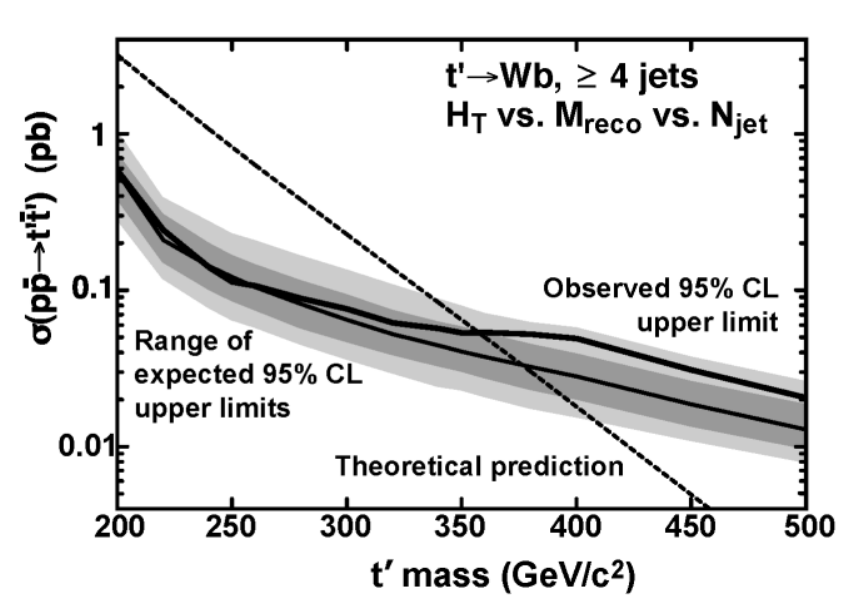}
\caption{Observed and expected 95\% C.L. upper limits as a
function of the mass of the $t'$ quark, for a $t'$ decaying to $Wb$ using 5.6~\invfb\ of CDF data~\cite{CDF_tprim}.}
\label{fig:CDF_tprim}
\end{figure}

In another search for fourth-generation quark based on 5.6~\invfb, CDF assumes that the new quark is heavier than the top, and for the purpose of setting limits also assumes that the decay chain is identical to the one of the top quark with the same couplings~\cite{CDF_tprim}. The selection requires $b$-tagging and the \tprim\ mass is reconstructed similarly to what is done for the top quark mass measurement using template methods. A two dimensional fit of the observed visible transverse energy where the missing transverse energy in the event is included and of the reconstructed mass distribution is performed to discriminate the new physics signal from standard model backgrounds. CDF excludes a standard model fourth-generation \tprim\ quark with mass below 358~\GeV\ with this analysis, as displayed in Fig.~\ref{fig:CDF_tprim}.

\begin{figure}
\centering
\includegraphics[width=.9\linewidth,height=.4\linewidth]{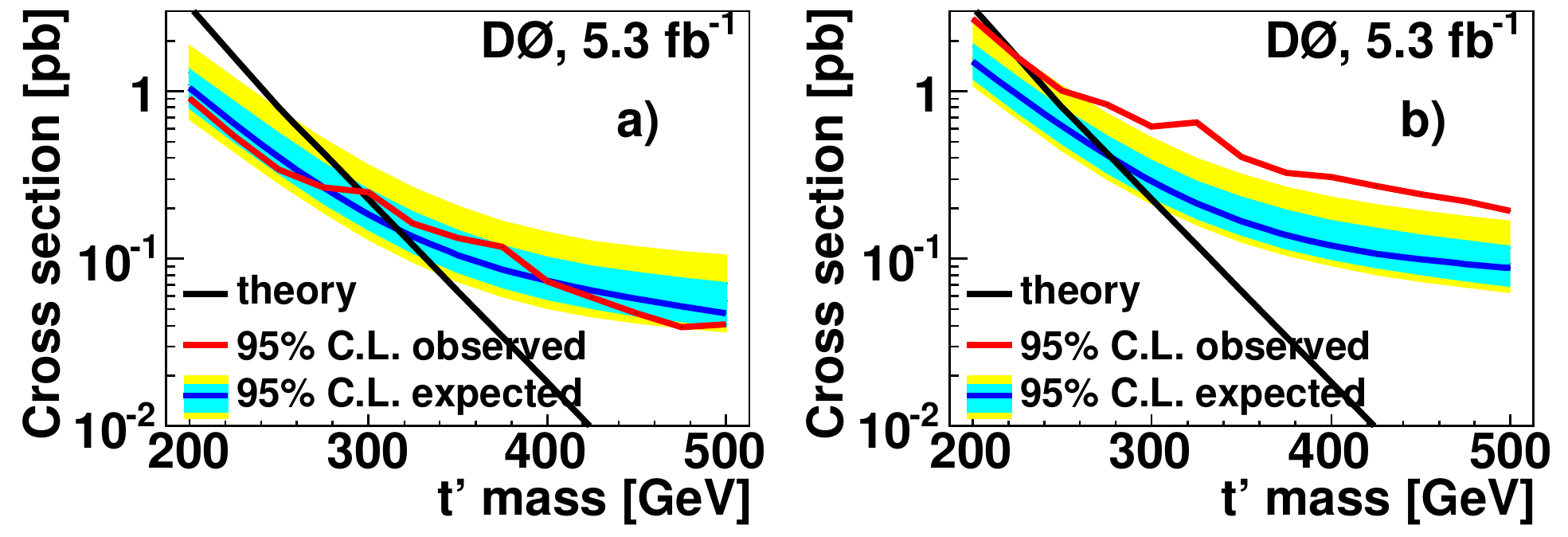}
\caption{Observed and expected upper limits and predicted values for the $t'\bar{t'}$ production cross section by the D\O\ search as a function of the mass of the $t'$ quark for (a) $e$+jets and (b) $\mu$+jets~\cite{D0_tprim}.}
\label{fig:D0_tprim_Wq}
\end{figure}

The D\O\ analysis based on 5.3~\invfb\ does not make explicit flavor tagging for the quark which can be any of the standard model downtype quarks~\cite{D0_tprim}. Similarly to the previous analysis, a 2D fit is performed. The $e$+jets and $\mu$+jets data are analyzed separately. There is no visible excess in the $e$+jets data.
In the $\mu$+jets data D\O\ observes a small excess of events over standard model expectations. Figure~\ref{fig:D0_tprim_Wq} shows the resulting cross section limits compared to the expected limits. In the $\mu$+jets sample, one can fit the data best with a \tprim\tprim\ production cross section of $3.2\pm1.1$ times the theoretical cross section for a \tprim\ quark mass of 325~\GeV. Due to this excess the lower mass limit is 285~\GeV\ for an expected sensitivity at 320~\GeV\ for $e+\mu$ combined.

A search for a  $4^{\rm th}$ generation $T^{\prime}$ in the decay mode $ t + X$, where $X$ is an invisible particle which could be dark matter has also been presented by CDF~\cite{CDF_tprimX}.
Here again the \ttbar\ lepton+jet sample is tested. There are several control regions defined to validate the modeling of the standard model background. For instance a control region is defined by requiring the missing transverse energy to be greater than 100~\GeV\ and requiring exactly 3 jets which is depleted of signal events. After applying a fitting template of the signal and background shapes to the observed events in the data using the $W$ transverse mass variable, one can excludes a $T^{\prime}$ as function of the mass of the invisible particle as shown in Fig.~\ref{fig:CDF_tprim_X}. It is also possible to reinterpret the cross section limits in terms of the supersymmetry decays of the stop into a top and a neutralino, although that one cannot make any mass exclusion for this decay mode.

\subsection{\ttbar\ resonances}
\begin{figure}
\centering
\includegraphics[width=.5\linewidth,height=.5\linewidth]{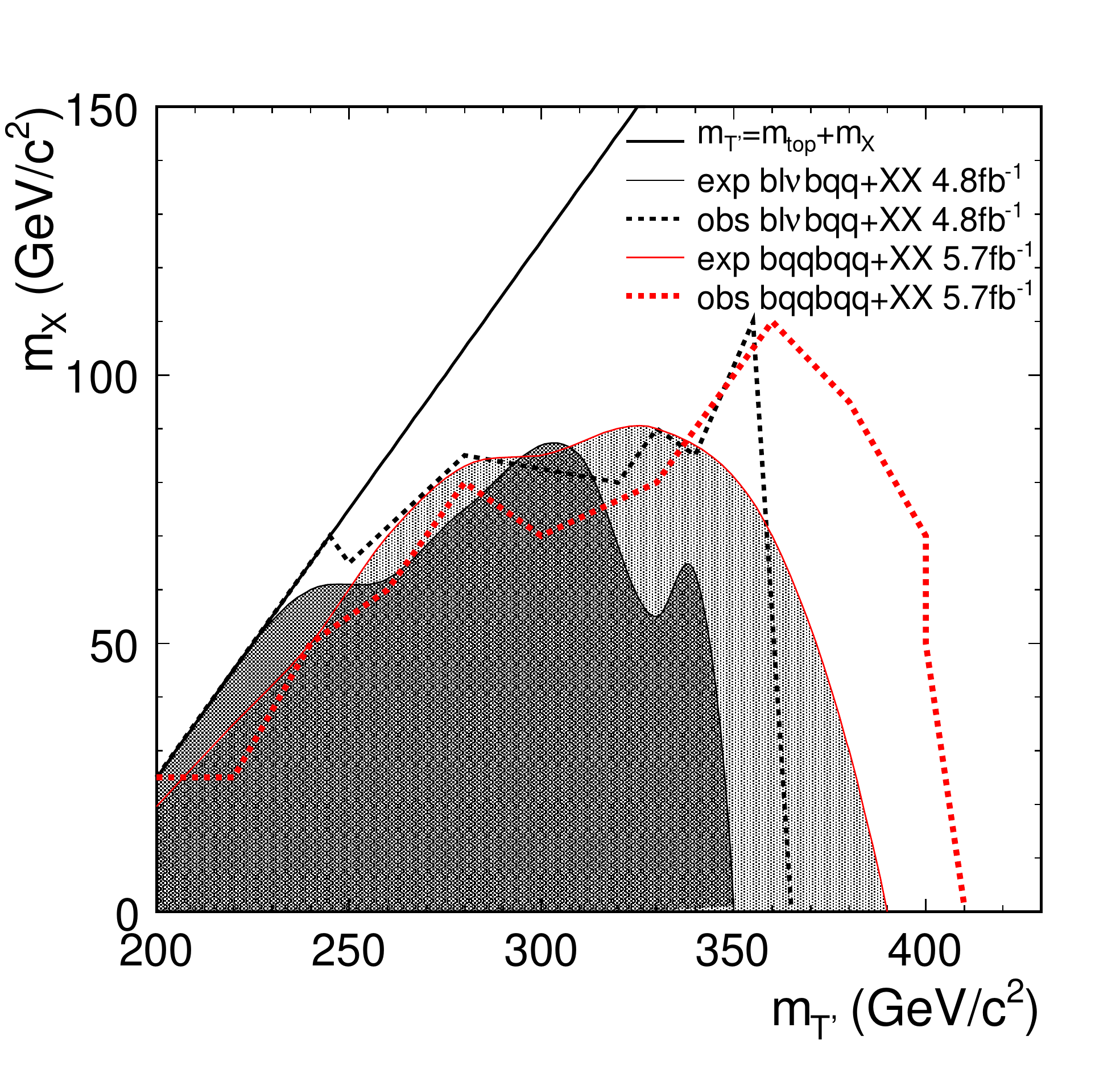}
\includegraphics[width=.45\linewidth,height=.45\linewidth]{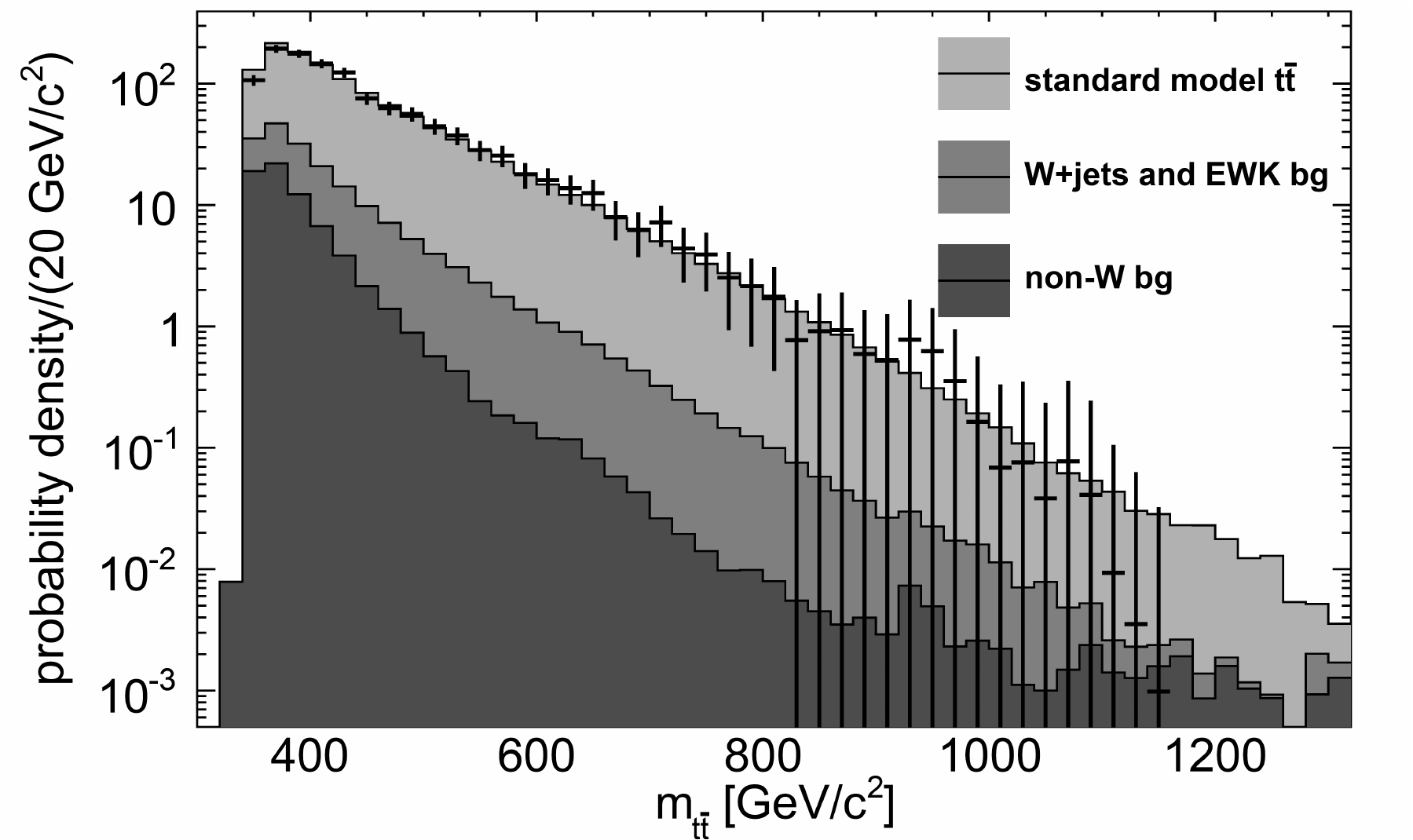}
\caption{The figure on the left shows the expected (exp) and observed (obs) 95\% C.L. exclusion region by CDF in the ($m_{T^{\prime}}, m_{X}$) parameters space where an exotic fourth generation quark $T^{\prime}$ decays to a top quark and an invisible particle~\cite{CDF_tprimX}. The figure on the right displays the $\ttbar$ invariant mass probability density function for each event used in the CDF search for resonant production of $t\bar{t}$ pairs in 4.8~\invfb\ of data~\cite{ref:CDF_ttreso}.}
\label{fig:CDF_tprim_X}
\end{figure}

It is also interesting to search for explicit \ttbar\ resonances because it is sensitive to new heavy neutral gauge bosons, Kaluza-Klein graviton excitations, axigluons or strong dynamics to name a few.
However, to test these models one should assume a narrow resonance to be able to detect the signal over the continuum \ttbar\ pair production background as shown in Fig.~\ref{fig:CDF_tprim_X} by the CDF search~\cite{ref:CDF_ttreso}. In this search, there is no evidence of \ttbar\ resonance found in 4.8~\invfb\ and cross section limits are compared to the cross section predicted for a leptophobic Topcolor \Zprim\ boson, allowing to set a lower mass limit at 900~\GeV.

Table~\ref{tab:summary_4gen} summarizes the Tevatron mass limits at the 95\% C.L. for $4^{\rm th}$ generation of fermion.

\section{No stone left unturned ?}
So far we have looked at many possible signatures of new physics over the past 10 years and the main question at this point remains:
are we sure we have no stone left unturned? What about supersymmetric particles for instance?

Searches for supersymmetry ({\tt SUSY}) have been conducted at both Tevatron experiments in a large variety of models and now the LHC has further
constrained most of the standard {\tt SUSY} scenarios we have been looking for. Is there any hope left for supersymmetry in Tevatron's data?
Probably not: squarks and gluinos seem beyond the kinematic reach of the Tevatron, processes produced via the electroweak interaction (like charginos) seem statistically limited, and the astrophysical cold dark matter candidate (neutralino) remains invisible.
At this point, in our quest for {\tt SUSY}, we try to concentrate on unusual decays or signatures which could have been missed by standard searches.
For instance there are several models which predict long-lived particles. Some of them are within the framework of supersymmetry.

\section{Supersymmetry long-lived}
Long-lived models predict that such exotic particles can be charged or neutral, decaying inside our outside the detector leading to unusual signatures. D\O\ released a search for a charged long lived massive particle decaying outside the detector using 5.2~\invfb\ of data~\cite{ref:D0_cmsp}. In these {\tt SUSY} inspired models, the lightest chargino or the lightest scalar tau lepton (stau) could play the role of such long lived charged particles and be the next to lightest supersymmetric particle. They decay into the lightest supersymmetric particle called the {\tt LSP}, which is stable and neutral. The next-to-lightest supersymmetric particle ({\tt NLSP}) can be long lived due to weak coupling to the {\tt LSP}.
In {\tt GMSB} model, the {\tt NLSP} is the stau. Other scenarios assume that the chargino and the neutralino {\tt LSP} are nearly degenerate in mass and there are two general cases: the first one where the chargino is mostly higgsino and the second one where the chargino is mostly gaugino, which are treated separately in the analysis.

\begin{figure}
\centering
\includegraphics[width=.6\linewidth,height=.45\linewidth]{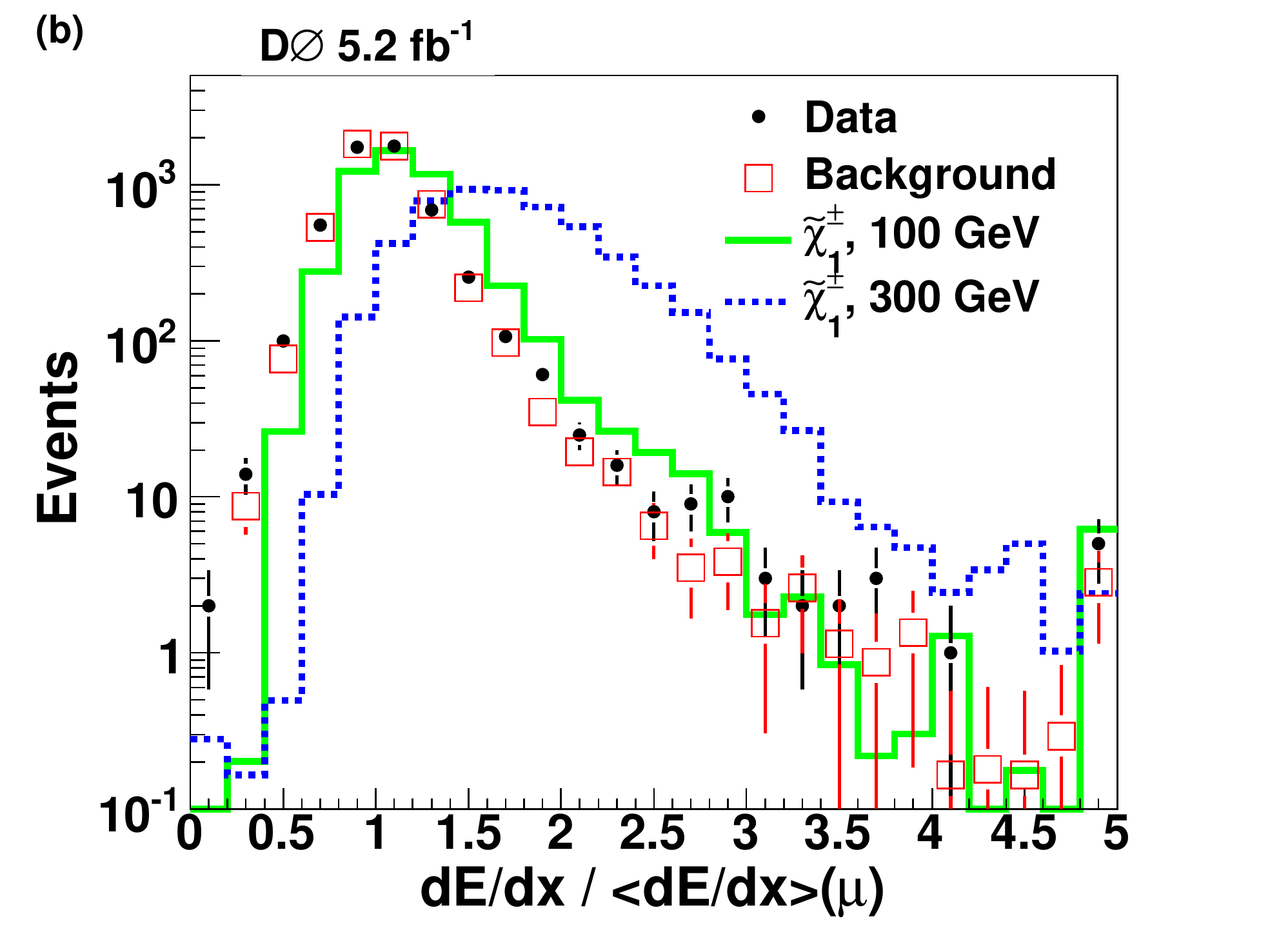}
\caption{Distribution of ionization energy loss ($dE/dx$) for data, background,
and signal (gaugino-like charginos with a mass of 100 and 300~\GeV) in the D\O\ search for charged massive long-lived particles~\cite{ref:D0_cmsp}. The scale is adjusted such that the muon $dE/dx$ measurement from \Zmumu\ events peaks at 1.}
\label{fig:D0_CMSP}
\end{figure}

These particles are stable, charged and massive. Therefore, this search makes use of the following two distinguishing experimental characteristics:
the speed ($\beta$) which is measured by muon scintillator counters, and the high $dE/dx$ as measured by the silicon detector and represented in Fig.~\ref{fig:D0_CMSP}. Dimuons events from decays of $Z$ bosons are used to calibrate the timing measurement and the $dE/dx$. Events from a $W$ decay with a mismeasured muon providing inaccurate values of the muon's speed and $dE/dx$ constitute the largest background and is evaluated in a control region defined by $\beta>1$ and a transverse mass of the $W$ system below 200~\GeV. This search is the first to extract $dE/dx$ from the D\O\ data.

D\O\ finds no evidence of signal above background and provide mass limits for particles predicted by several different models, the observed cross section limits are generally applicable to direct pair production of any charged massive long-lived particles ({\tt CMLLPs}) of a given mass.
Using this search, D\O\ is able to set only cross-section upper limits for a pair-produced staus model.
Exclusion mass is however possible for pair-produced long-lived stops with mass below 265~\GeV,
gaugino-like charginos below 251~\GeV, and higgsino-like charginos below 230~\GeV. These are presently the most restrictive limits for chargino {\tt CMLLPs}.

\section{Other exotic final states}

\subsection{Search for New Physics in ZZ+MET}
 As mentioned earlier, a $4^{\rm th}$ generation of particles would be a natural extension of the existing three families present in the standard model. What about having the lightest $4^{\rm th}$ generation to be a neutrino? The largest production mechanism could also be via a Drell-Yan process as $p\bar{p} \rightarrow Z/\gamma^{*} \rightarrow N_2 N_2 \rightarrow N_1 Z N_1 Z$ where $N_2$ is the unstable heavy eigenstate and $N_1$ is the stable and least massive eigenstate of the $4^{\rm th}$ generation neutrinos. CDF isolates this signature in 4~\invfb\ by requiring two charged leptons and two jets both coming from the $Z$ and large missing transverse momentum~\cite{ref:CDF_zzmet}. The difference in mass between the two reconstructed $Z$ has the signal peaks at low mass difference. CDF finds agreement between data and standard model backgrounds and set cross section limits of $\sim$~300~\fb\ and yields limits in the magnitude of 14 events across a spectrum of generated masses.

\begin{figure}
\centering
\includegraphics[width=.32\linewidth,height=.3\linewidth]{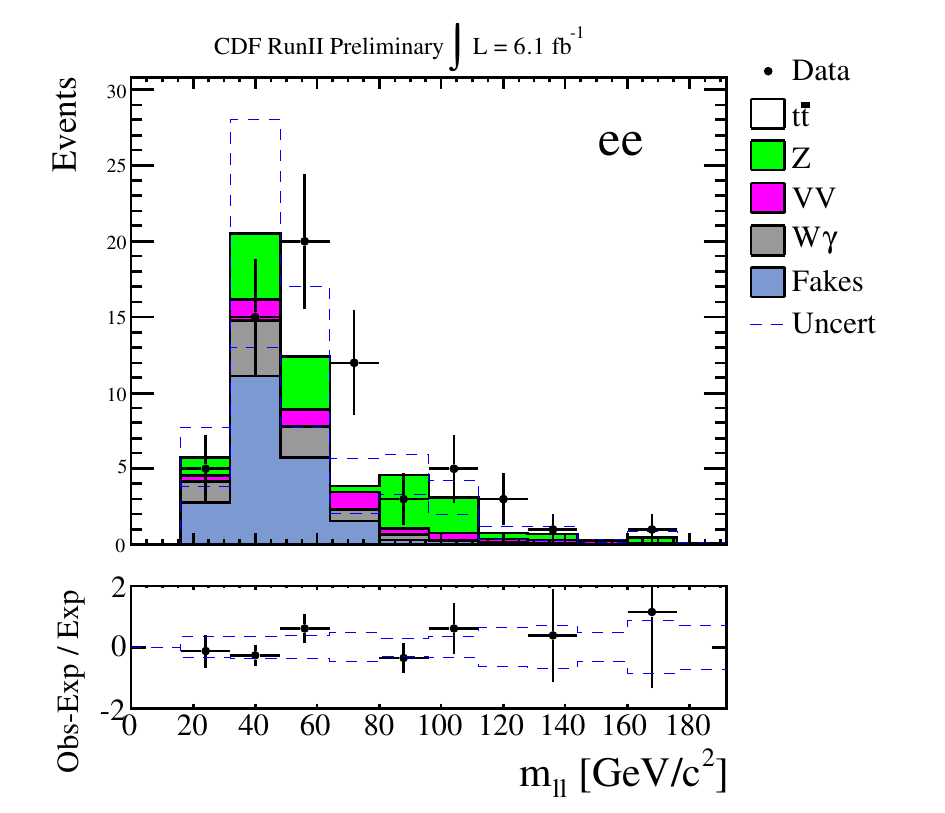}
\includegraphics[width=.32\linewidth,height=.3\linewidth]{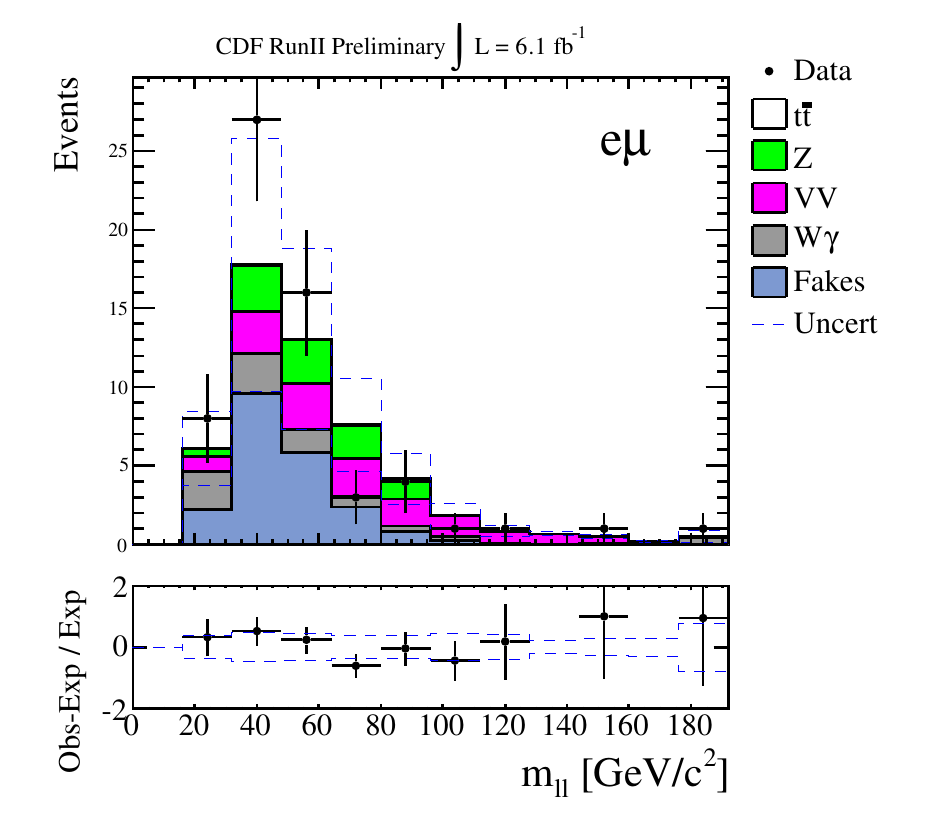}
\includegraphics[width=.32\linewidth,height=.3\linewidth]{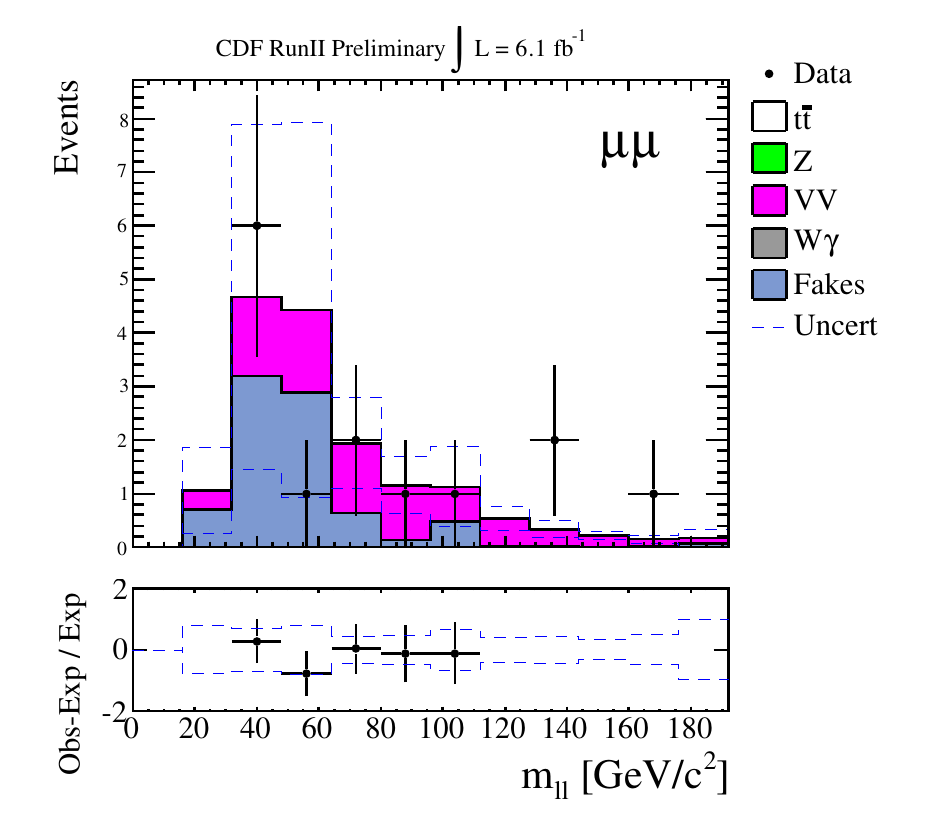}
\caption{Distribution of dilepton invariant mass in observed same-sign dilepton events ($ee$, $e\mu$, $\mu\mu$) as measured by CDF in 6.1~\invfb\ of data~\cite{ref:CDF_ss}.}
\label{fig:CDF_ss}
\end{figure}

\subsection{Same sign dilepton}
A wide variety of new physics models predict events with two like-sign leptons, a signature which has very low backgrounds from the standard model.
Examples include Universal Extra dimension, supersymmetry or same-sign top quark. CDF examines this final states in 6.1~\invfb\ of data~\cite{ref:CDF_ss}. The dominant background comes from events in which the second lepton is due to the semi-leptonic decay of a $b$- or $c$-quark meson, largely from $W$+jets production or \ttbar\ production. Such a background is described using a lepton misidentification model from inclusive jet data applied to $W$+jet events. Figure~\ref{fig:CDF_ss} shows the dilepton invariant mass distributions of observed and predicted same-sign lepton events. Limits on several simplified models are obtained since data are consistent with standard model expectations.

\section{Any other hints of new physics ?}
At that point, are there any other hints of new physics beyond the ones we have discussed so far?

\subsection{Multijet resonances in $bbb$}
\begin{figure}
\centering
\includegraphics[width=.25\linewidth,height=.25\linewidth]{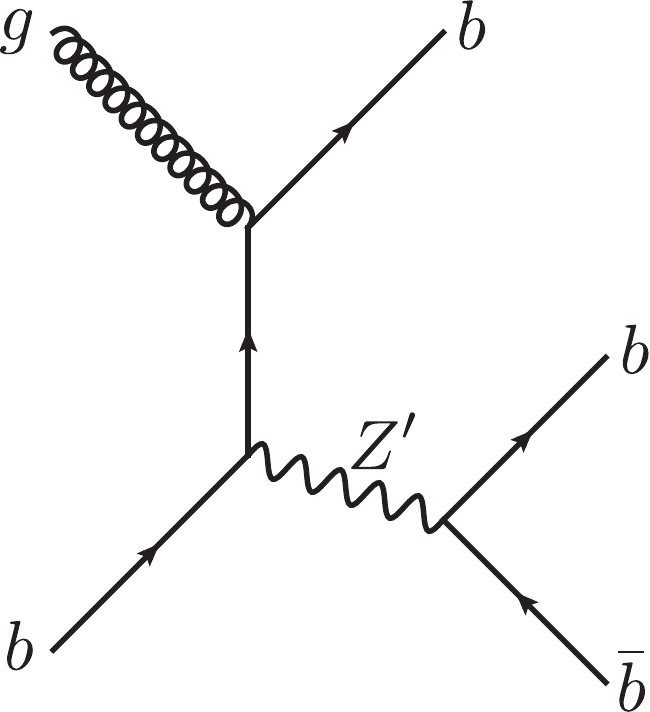}
\ \ \ \ \ \
\includegraphics[width=.4\linewidth,height=.3\linewidth]{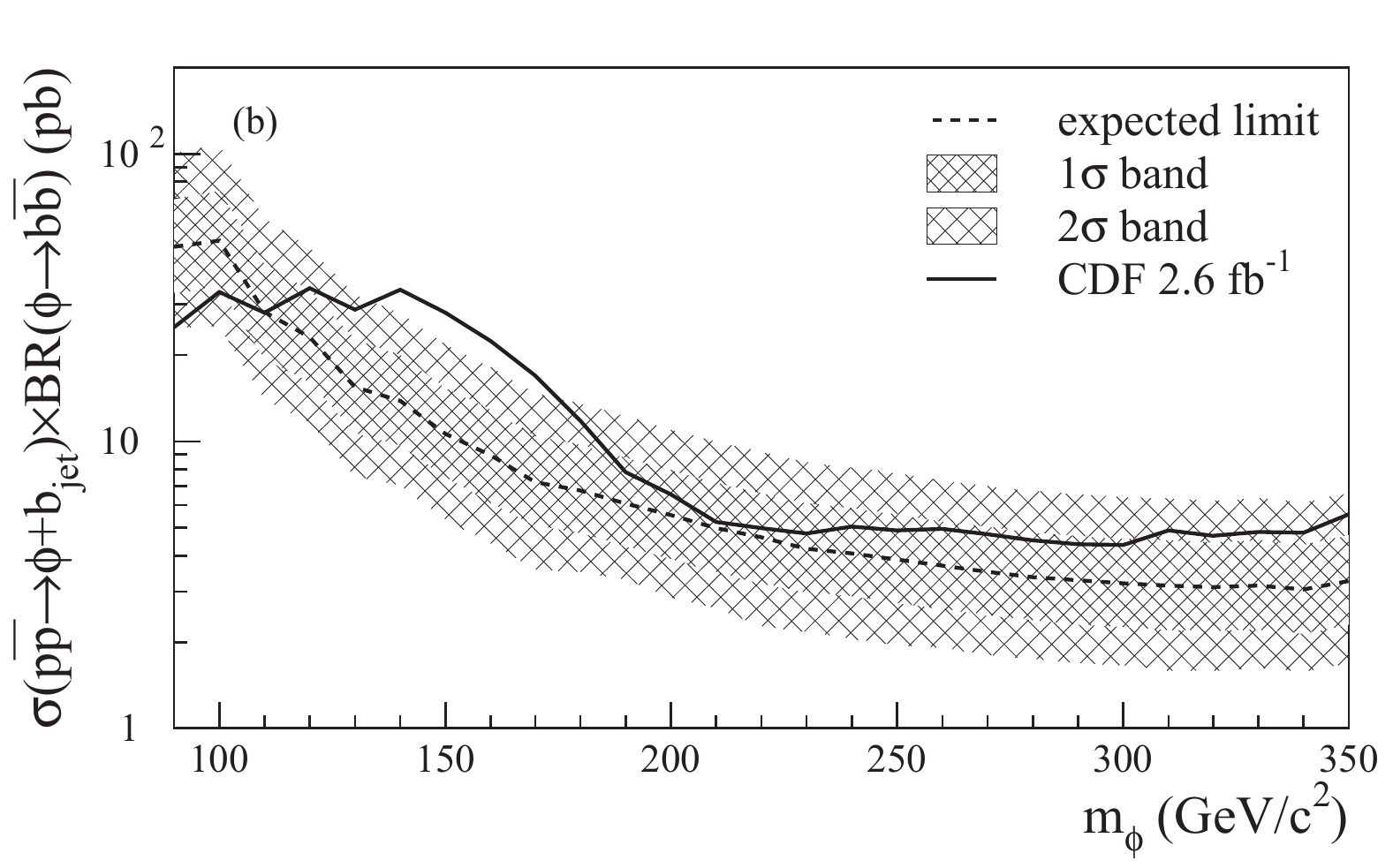}
\caption{({\it left}) Feynman diagram leading to final states with three b-jets through an on-shell \Zprim~\cite{ref:zprim_bb} ;  ({\it right}) Expected and observed cross section limits in the CDF search for a neutral Higgs bosons $\phi$ decaying into
$b\bar{b}$, produced in association with $b$ quarks in 2.6~\invfb\ of data~\cite{ref:CDF_bbb}.}
\label{fig:CDF_bbb}
\end{figure}

Based on 2.6~\invfb of data, CDF has reported a cross section limit that is weaker than expected for Higgs masses in the range of 130 to 160~\GeV\ in the context of the search for {\tt SUSY} Higgs decaying to \bb\ and in association with an additional $b$-jet~\cite{ref:CDF_bbb}. Including the trials factors to account for the number of mass points searched in steps of 10~\GeV\ in the mass range tested (90-350~\GeV), CDF expects to see a deviation of this size at any mass in 2.5\% of background only pseudoexperiments. However, given the cross-section required for such an excess, it is difficult to attribute this excesses to a Higgs boson. On the other hand, if we allow the \Zprim\ to have large couplings to $b$ quarks~\cite{ref:zprim_bb}, collisions at the Tevatron can lead to events with three $b$-jets as illustrated by the Feynman diagram shown in Fig.~\ref{fig:CDF_bbb}. This could be responsible for the excess of multi-b events. The expected and observed limits are plotted as a function of the narrow scalar mass in Fig.~\ref{fig:CDF_bbb}.

\subsection{Top anti-Top asymmetry}
The forward-backward asymmetry in top quark pair production is another source of hint of new physics reported at the Tevatron~\cite{ref:cdfasymmetry}.
There are numerous theoretical papers suggesting interesting new physics mechanism including axigluons, diquarks, new weak bosons or extra dimension to explain this anomaly.  One possible explanation for this discrepancy could be a flavor-violating chirally \Zprim\ boson that mixes to up and top quarks where contributions in the t-channel will not necessarily lead to unacceptable modification of the \ttbar\ cross section~\cite{ref:zprim_bb}. However, such observed asymmetry could also be very well due to the modeling (by generators) of gluon radiations at {\tt NLO} rather than to a disagreement with the standard model itself. The on going more precise measurements with the full Tevatron dataset will therefore be very interesting to follow. The proceeding~\cite{ref:FDeliot} related to the top quark at the Tevatron gives more details about the current available measurements including a new result from D\O~\cite{ref:d0asymmetry}.

\subsection{Dimuon Asymmetry and $B_s$ measurement}
There are also indirect hints coming from the $B$ physics sector that were discussed in the flavor physics session and that are described in detail in the corresponding proceeding~\cite{ref:DTonelli}.

The first hint is the update measurement from D\O\ of the anomalous like sign dimuon charge asymmetry where the notable 1\% excess of the matter particles is confirmed at 3.9 standard deviation level using a dataset of 9~\invfb\ providing evidence for large CP violation beyond what is predicted by the standard model~\cite{ref:D0dimuasym}. The dependence of the asymmetry on the muon impact parameter is consistent with the hypothesis that it originates from semileptonic b-hadron decays.

Finally, a new measurement was released by CDF of the Branching ratio of $B_s$ into dimuon with an excess ($<3 \sigma$) of events in the search windows using 6.9~\invfb\ of data~\cite{ref:CDF_Bs}. At the conference, LHCb also released an analysis of the data recorded in the first half of 2011 corresponding to an integrated luminosity of 0.37~\invfb~\cite{ref:LHCbmumu}. The search for \bsmm\ by LHCb do not confirm the excess seen by CDF. In susy for instance, enhancement of $\tan \beta$ is expected making the study of these decays among the most sensitive for indirect searches for new physics.

\section{Conclusion}
A huge number of signatures were explored at the Tevatron with up to 9~\invfb\ of data analyzed so far. A few hints are summarized in the table~\ref{tab:summary_TevHints}. Most of them are being followed up on which makes the Tevatron dataset (11~\invfb) still very interesting. We are in a new era of large data samples both at the Tevatron and the LHC. In the absence of a discovery at the Tevatron, our goal is to leave no stone unturned.

\acknowledgments
I would like to thank my colleagues both at Fermilab and at other collaborating institutions, especially those who operated
the Tevatron accelerator and constructed, maintained, and calibrated the CDF and D\O\ detectors, essential for any physics analysis
reported here. Special thanks to the Fermilab Accelerator Division for providing proton-antiproton collisions with remarkable reliability at beam intensities far beyond the design goals of the Tevatron and to recognize those who have helped make the CDF and D\O\ experiments a success over the years. This includes computing divisions and funding agencies for making all of this possible. Thanks also to our families for their everyday support and during the number of late nights and long weekends also making all of this possible.

Many thanks to the local organizing committee chaired by Johann Collot for this very well organized conference. Finally, my gratitude also extends to all members of the European Physical Society's High Energy Particle Physics Division Board of EPS-HEP2011 for inviting me to give this plenary talk and special thanks to Fabio Zwirner and Paris Sphicas for the discussions during the preparation of this talk. It was a great honor and a pleasure for me to have been given such an opportunity to talk about new physics at the Tevatron.

\begin{center}
\begin{table}[htbp]
\begin{center}
\begin{tabular}[!htb]{|c|c|c|c|c|c|c|c|c|c|}
\hline
                & \multicolumn{9}{c|}{\bf Sequential V'}\\[3pt] \hline
 X $\raa$       & $e\nu$ & $ee$ & $\mu\mu$ & $e\mu$ & $\tau\tau$ & $qq$ & $tt$	& $tb$ & $WZ$ \\[3pt] \hline
 Limits (\GeV)	& 1120	& 1023  & 1071  &  700	& 399  & 740 & 900 & 885 & 180-690 \\[3pt] \hline
Experiment      &  CDF  & D\O   &  CDF  &  CDF  &  CDF & CDF & CDF & D\O    &  D\O \\[3pt]
[Ref.]          & \cite{ref:CDF_W_emu} & \cite{ref:D0_Zprim_ee} & \cite{ref:CDF_Zprim_mumu} & \cite{ref:CDF_Zprim_emu} & \cite{ref:CDF_Zprim_tautau} & \cite{ref:CDF_Zprim_qq} & \cite{ref:CDF_ttreso} & \cite{ref:D0_Wprim_tb} & \cite{ref:D0_reso_WW_ZZ} \\[3pt] \hline
\end{tabular}
\end{center}
\caption{Summary of the Tevatron mass limits at the 95\% C.L. in the context of searches for a resonance using
the sequential standard model \Wprim\ ($e\nu$, $WZ$) and \Zprim\ ($ee$, $\mu\mu$, $\tau\tau$, $qq$) bosons, purely right-handed
couplings \Wprim\ ($tb$) boson, leptophobic \Zprim\ ($tt$) boson, and lepton family
number violating $E_6$-like model of $U(1)^{'}$ ($e\mu$). Only the most stringent limits are quoted.}
\label{tab:summary_Vprim}
\end{table}
\begin{table}[htbp]
\begin{center}
\begin{tabular}[!htb]{|c|c|c|c|c|c|c|c|}
\hline
                & \multicolumn{7}{c|}{\bf RS-G. $k/M_{\hbox{\rm Pl}}=0.1$}\\[3pt] \hline
X $\raa$        & $ee$ & $\gamma\gamma$ & $\mu\mu$ & $ee+\gamma\gamma$ & $ee+\mu\mu+\gamma\gamma$ & $WW$ & $ZZ$  \\[3pt] \hline
Limits (\GeV)	& 914	& 963 &	859	& 1058 & 1111 &	300-754	& 600 \\[3pt] \hline
Experiment      &  CDF  & CDF & CDF &  CDF &  CDF & D\O    &  CDF \\[3pt]
[Ref.]          & \cite{ref:CDF_ZprimRS_ggee} & \cite{ref:CDF_RS_gg} & \cite{ref:CDF_RS_mumueegg}  & \cite{ref:CDF_ZprimRS_ggee} & \cite{ref:CDF_RS_mumueegg}  & \cite{ref:D0_reso_WW_ZZ} & \cite{ref:CDF_ZZ_4l_6fb} \\[3pt] \hline
\end{tabular}
\end{center}
\caption{Summary of the Tevatron mass limits at the 95\% C.L. in the context of searches for a resonance using the Randall-Sundrum model graviton (RS-G.). Only the most stringent limits are quoted.}
\label{tab:summary_RS}
\end{table}
\begin{table}[p]
\begin{center}
\begin{tabular}[!htb]{|c|c|c|c|c|c|c|c|}
\hline
                & \multicolumn{4}{c|}{\bf $4^{\rm th}$ generation} & \multicolumn{2}{c|}{\bf Vector quarks} \\[3pt] \hline
Process         & $\bprim \raa tW$ & $\tprim \raa qW $ & $\tprim \raa bW$ & $\tprim \raa tX$ & $qW$ & $qZ$ \\[3pt] \hline
Limits (\GeV)	&              372 &	285          & 358              & 360               & 693 &	551  \\[3pt] \hline
Experiment      & CDF              & D\O             & CDF              & CDF               & D\O & D\O \\[3pt]
[Ref.]          & \cite{CDF_bprim} & \cite{D0_tprim} & \cite{CDF_tprim} & \cite{CDF_tprimX} & \cite{ref:D0_VQ} & \cite{ref:D0_VQ} \\[3pt] \hline
\end{tabular}
\end{center}
\caption{Summary of the Tevatron mass limits at the 95\% C.L. for $4^{\rm th}$ generation of fermion and vector-like quarks. Only the most stringent limits are quoted.}
\label{tab:summary_4gen}
\end{table}

\begin{table}[htbp]
\begin{center}
\begin{tabular}[!htb]{|c|cc|c|c|c|}
\hline
{\bf Process}       & {\bf Exp.}& {\bf [Ref.]} & {\bf Luminosity} & {\bf Significance} & {\bf Mass}  \\[3pt] \hline
$ZZ \raa$ 4 leptons & CDF & \cite{ref:CDF_ZZ_4l_6fb} & 6~\invfb & 3$\sigma$> & 325~\GeV \\[3pt] \hline
$W+jj$              & CDF & \cite{CDF_Wjj_73_bump} & 7.3~\invfb & 2-4.1$\sigma$ & 150~\GeV \\[3pt] \hline
$\tprim \raa qW $   & D\O & \cite{D0_tprim} & 5.3~\invfb & 2.5$\sigma$ & 325~\GeV \\[3pt] \hline
$bbb$               & CDF & \cite{ref:CDF_bbb}& 2.6~\invfb & 2.8$\sigma$ & 150~\GeV \\[3pt] \hline
\ttbar\ Asym.       & CDF & \cite{ref:cdfasymmetry}& 5.1~\invfb & 3.4$\sigma$ &  \\[3pt] \hline
\ttbar\ Asym.       & D\O & \cite{ref:d0asymmetry}& 5.4~\invfb & 2.4$\sigma$> &  \\[3pt] \hline
$\bsmm$             & CDF & \cite{ref:CDF_Bs}& 5.1~\invfb & 3$\sigma$< &  \\[3pt] \hline
dimuon Asym.        & D\O & \cite{ref:D0dimuasym}& 9.0~\invfb & 3.9$\sigma$ &  \\[3pt] \hline
\end{tabular}
\end{center}
\caption{Summary of Tevatron's hints in the data. The probability that the yield observed in data results from a statistical
fluctuation of the standard model sample is converted into standard deviations for each analysis. Note that the correction
factor for the number of trials is not included. The invariant mass of clustered events is given in the event of a resonance.}
\label{tab:summary_TevHints}
\end{table}
\end{center}


\end{document}